\documentclass[aps,twocolumn,superscriptaddress,nofootinbib,floatfix]{revtex4-2}

\usepackage{array}[=2016-10-06]
\usepackage[utf8]{inputenc}
\usepackage{graphicx,bm,bbm,amsmath,booktabs,tabularx,amssymb,amsthm,mathtools}
\usepackage{xcolor}
\usepackage{physics}
\usepackage{graphicx}
\usepackage{float}                 
\usepackage[caption=false]{subfig}
\usepackage[colorlinks,citecolor=blue,linkcolor=blue,urlcolor=blue]{hyperref}
\newcommand{\bw}{{\bm{w}}}
\newcommand{\bx}{{\bm{x}}}
\newcommand{\ii}{\mathrm{i}}
\newcommand{\E}{\mathbb{E}}
\newcommand{\PP}{\mathbb{P}}
\newcommand{\de}{\mathrm{d}}
\newcommand{\sgn}{\operatorname{sgn}}
\newcommand{\Dt}{\mathcal{D}t}
\newcommand{\spin}{{\rm sp}}
\newcommand{\bW}{\mathbf{W}}
\newcommand{\bh}{\bm{h}}
\newcommand{\R}{\mathbbm{R}}
\newcommand{\Z}{\mathbbm{Z}}
\newcommand{\one}{\mathbf{1}}
\newcommand{\asympeq}{\mathrel{\overset{\mathrm{exp}}{=}}}
\newcolumntype{Y}{>{\raggedright\arraybackslash}X}

\newtheorem{prop}{Proposition}

\newtheorem{thm}{Theorem}
\theoremstyle{remark}
\newtheorem{remark}{Remark}

\usepackage{soul}

\begin{document}

\title{Periodic Readout Induces a First-Order Generalization Transition in a Quantum Perceptron}

\author{Fabio Benatti}
\affiliation{Department of Physics, University of Trieste, 34151 Trieste, Italy}
\affiliation{INFN Sezione di Trieste, 34151 Trieste, Italy}
\author{Masoud Gharahi}
\affiliation{Faculty of Physics, Astronomy and Applied Computer Science, Jagiellonian University, 30-348 Krak\'ow, Poland}
\author{Giovanni Gramegna}
\affiliation{Department of Physics, University of Bari, 70126 Bari, Italy}
\affiliation{INFN Sezione di Bari, 70126 Bari, Italy}
\author{Stefano Mancini}
\affiliation{School of Science and Technology, University of Camerino, 62032 Camerino, Italy}
\affiliation{INFN Sezione di Perugia, 06123 Perugia, Italy}
\author{Vincenzo Parisi}
\affiliation{CONCEPT Lab, Fondazione Istituto Italiano di Tecnologia, 16152 Genova, Italy}

\date{\today}

\begin{abstract}
Variational quantum classifiers can generate highly structured decision boundaries through the Fourier dependence of observable expectation values on circuit parameters. We show that expressivity can come at a sharp statistical cost, by considering a quantum perceptron whose readout is the simplest periodic classifier, namely a square-wave function of the weighted inputs, and determine its typical generalization in the thermodynamic limit using a teacher-student replica-symmetric theory. Above a critical oscillation frequency, increasing the training load induces a first-order transition from weakly to strongly teacher-aligned phases, resulting in a discontinuous improvement in generalization. At fixed load, however, increasing the frequency eventually drives the equilibrium generalization error to chance level, despite perfect classification of all the training examples. Thus, the same periodic structure that enhances random-label storage can delay the onset of learning. Our results expose a quantitative tradeoff between memorization and generalization generated purely by periodic readout, and provide a benchmark for understanding Fourier-rich variational quantum classifiers.
\end{abstract}

\maketitle

\section{Introduction}
Parametrized quantum circuits are the workhorse of near-term quantum machine learning~\cite{biamonte2017,Cerezo2021,Benedetti2019,Mitarai2018,FarhiNeven2018}. Their predictions are expectation values of quantum observables on the output of the circuit, which admit a Fourier representation in terms of trigonometric polynomials in the rotation angles used in the circuit, with frequencies set by the spectra of the gate generators~\cite{SchuldSwekeMeyer2021,PerezSalinas2020}. This oscillatory response to the trainable parameters, a direct consequence of unitary rotations followed by a projective measurement, is usually regarded as a source of expressivity. 
{For the capacity to \emph{store} random input-label associations, the role of the oscillation is known for a minimal quantum perceptron: a higher readout oscillation frequency $\lambda$ increases the storage capacity~\cite{BGGMP26}.}
Its effect on \emph{learning} from data is much less understood: generalization bounds for quantum models~\cite{Caro2022,Banchi2021,GilFuster2024} are worst-case statements, while the statistical mechanics of learning has shown that typical learning curves of even the simplest networks can be qualitatively different~\cite{Gyorgyi1990,Sompolinsky1990,Seung1992,watkin1993,engel2001}.

{Here we address the question of \emph{rule inference},} in the teacher-student scenario in which a student perceptron must reproduce the labels of a teacher of identical architecture from $P=\alpha N$ examples, where $N$ is the dimensionality of the data. We find that the same frequency that enlarges the storage capacity has the opposite effect on learning: {(i) the initial alignment with the teacher is exponentially suppressed in $\lambda^2$;} (ii) for $\lambda>\lambda_c\simeq1.944$ learning occurs through a first-order transition; {(iii) the transition load grows only logarithmically with $\lambda$, whereas a weakly aligned metastable phase survives up to loads $\sim {\rm e}^{\lambda^2}/\lambda^4$, opening a wide region that is expected to be hard for efficient inference algorithms; (iv) the asymptotic learning curve is nonetheless $\lambda$ independent.}

\section{Model}
\begin{figure}[t]
\includegraphics[width=\columnwidth]{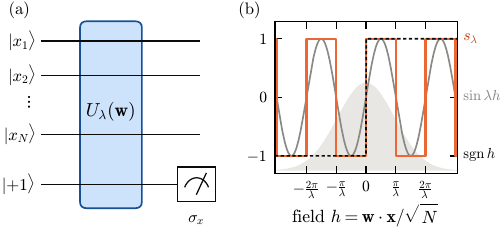}
\caption{(a) Quantum perceptron: basis-encoded input, trainable gate~\eqref{eq:U}, and $\sigma_x$ readout. (b) Readout~\eqref{eq:readout} (gray), square-wave classifier $s_\lambda$ (orange, $\lambda=3$), classical perceptron $\sgn(h)$ (dashed black), and the Gaussian distribution of the fields $h$ (shaded).}
\label{fig:model}
\end{figure}

The input register consists of $N$ qubits prepared in $|\bx\rangle=|x_1\rangle\otimes\cdots\otimes|x_N\rangle$, where $\bx\in\{\pm1\}^N$, so that $\sigma_z^{(j)}|\bx\rangle=x_j|\bx\rangle$. The readout qubit is prepared in $|+1\rangle$; see Fig.~\ref{fig:model}(a). The trainable gate
\begin{equation}
U_\lambda(\bw)=\exp\Bigl(-\frac{\ii}{2}\,\lambda\,\hat{\bw}\cdot\bm{\sigma}_z\otimes\sigma_y^{(\rm out)}\Bigr),\qquad \hat{\bw}=\frac{\bw}{\|\bw\|},
\label{eq:U}
\end{equation}
acts on the input and readout register, jointly prepared in $\ketbra{\bm{x}}{\bx}\otimes \ketbra{+1}{+1}$, as a rotation of the readout register by the angle $\lambda\,\hat{\bw}\cdot\bx$. Then, the expectation value of measuring 
$\sigma_x^{(\rm out)}$ is~(Appendix~\ref{sec:circuit})
\begin{align}
f_\lambda(\bx,\bw)\!&\coloneqq\!\Tr[\!(U_\lambda(\bm{w})
\ketbra{\bm{x}}{\bx}\!\otimes\!\ketbra{+1}{+1}
U_\lambda(\bm{w})^\dagger)
\mathbb{I}\!\otimes\!\sigma_x^{(\rm out)}\!]\notag\\
&=\sin\Bigl(\lambda\,\frac{\bw\cdot\bx}{\|\bw\|}\Bigr).
\label{eq:readout}
\end{align}
Since the readout depends only on the direction of $\bw$, we fix the redundant radial degree of freedom by imposing $\|\bw\|^2=N$. For delocalized weights and independent Rademacher inputs, the local field $h=\bw\cdot\bx/\sqrt N$ is of order one. We define the classifier 
\begin{equation}\label{eq_sin}
s_\lambda(h)\coloneqq\sgn\bigl(\sin(\lambda h)\bigr)
\end{equation}
in terms of the local field $h=\bw\cdot\bx/\sqrt N$. The decision boundaries are given by the parallel hyperplanes $h=k\pi/\lambda$, $k\in\mathbb Z$, so that $\lambda$ fixes their density as seen by the $O(1)$ fields (see Fig.~\ref{fig:model}(b)), and $\lambda\to0^+$ recovers the classical perceptron $\sgn(h)$. We assume the ability to estimate  expectation values with no errors, i.e., we neglect shot noise associated to finite-sampling. Note that the circuit does not generate entanglement on basis inputs and is classically simulable: The effects investigated below therefore stem solely from the periodic structure of the quantum readout, not from some inherently quantum property.

\section{Teacher-student setting}
We formulate learning in the teacher--student framework, which provides
a controlled setting for studying how examples constrain an unknown
classification rule~\cite{watkin1993,engel2001}. A teacher generates the
training labels, and the student has access to the labeled examples
but not to the teacher's parameters. The question is how much these
examples reveal about the underlying rule, and whether this information
allows the student to predict labels on previously unseen inputs.

In our model, the teacher weight vector $\bw^*$ is drawn uniformly from the sphere $\|\bw^*\|^2=N$. Independently of the teacher, we draw $P$ inputs $\bx^\mu$ whose entries are independent and take the values $\pm1$ with equal probability. The corresponding labels are generated according to
\begin{equation}
\xi^\mu=s_\lambda\Bigl(\frac{\bw^*\cdot\bx^\mu}{\sqrt N}\Bigr),
\qquad \mu=1,\ldots,P.
\end{equation}
We write $X=(\bx^1,\ldots,\bx^P)$ and $\bm\xi=(\xi^1,\ldots,\xi^P)$ for the training inputs and labels and with $\mathcal{T}_P=\{(\bx^\mu,\xi^\mu)\}_{\mu=1}^P$ the whole training set.

For a given training set, every weight vector that reproduces all
observed labels remains a possible explanation of the data. These
compatible weights form the \emph{version space} of the student, that is assumed to use the same classification rule and the same spherical
prior as the teacher. Thus:
\begin{equation}
\mathcal{V}(\mathcal{T}_P)=\Big\{\bw\mid \,\|\bw\|^2=N,\
s_\lambda\Bigl(\frac{\bw\cdot\bx^\mu}{\sqrt N}\Bigr)
=\xi^\mu \quad \text{for all } \mu\Big\}.
\end{equation}
An incompatible weight vector has
zero likelihood, while every compatible vector has likelihood one.
Bayes' rule therefore restricts
the prior to the version space and normalizes the resulting measure.
If $\mathrm d\mu(\bw)$ denotes the normalized uniform measure on the
sphere, the posterior distribution is (Appendix \ref{sec:setting})
\begin{equation}
\de\mu_{\mathcal{T}_P}(\bw)=
\frac{\mathbf 1_{\mathcal{V}(\mathcal{T}_P)}(\bw)}{V_N(\mathcal{T}_P)}
\,\mathrm d\mu(\bw),
\end{equation}
where $\mathbf 1_{A}$ denotes the indicator function of the set $A$ ($\mathbf 1_{A}(\bw)=1$ if $\bw\in A$ and zero otherwise) and 
\begin{equation}
V_N(\mathcal{T}_P)=
\int_{\mathcal{V}(\mathcal{T}_P)}\mathrm d\mu(\bw)
\end{equation}
is the volume of the version space, also called \emph{Gardner volume}: The fraction of the original weight space
that remains compatible with the training
set~\cite{gard1987,gard1988,gard1988s,GD89}.
Thus, sampling the posterior amounts to drawing a student uniformly
from the compatible weights; we refer to such a posterior sample as
a \emph{Gibbs student}. 

The prior and labeling rule used for inference match those that
generated the data, which is the setting commonly called
\emph{Bayes-optimal inference}~\cite{Zdeborova2016}.
Its connection to statistical mechanics is expressed by the
\emph{Nishimori condition}, often described as being on the Nishimori
line~\cite{nish2001,Iba1999}. The relevant consequence here can be
understood directly from Bayes' rule. That is, conditioned on the observed
training set, the unknown teacher has exactly the same distribution
as a posterior-sampled student. The teacher and independent posterior
samples can therefore be exchanged in statistical averages. This
property underlies the Nishimori identities used below.

The Gardner volume also has a direct probabilistic interpretation.
Before the teacher weights are known, the probability of observing
the complete label sequence $\bm\xi$ at fixed inputs $X$ is obtained
by integrating the teacher prior over all weights that produce those
labels. Consequently, {$V_N(\mathcal T_P)=\PP(\bm\xi\,|\,X)$}.
In Bayesian terminology, this probability is the \emph{evidence} 
(Appendix \ref{sec:setting}).
The same quantity therefore measures both the fraction of compatible
weights and the probability assigned to the observed labels.

To characterize its behavior at large $N$, we consider the
\emph{quenched entropy density}, obtained by computing the logarithm
of the volume for each fixed training set and then averaging over the
teacher and inputs.  The evidence interpretation gives
$\mathbb{E}[\ln V_N]=-H(\bm\xi\,|\, X)$, the conditional
{(Shannon) entropy of $\bm\xi$ given $X$}. Moreover, the labels are completely determined by
the teacher and inputs, so
$H(\bm\xi\, | \,\bw^*,X)=0$. Hence, the conditional entropy
$H(\bm\xi\, | \, X)$ equals the conditional mutual information 
$I(\bw^*;\bm\xi\, | \, X)$, and, in the limit
$N,P\to\infty$ at fixed training load $\alpha=P/N$,
\begin{equation}
\begin{aligned}
\phi(\alpha,\lambda)
&{\coloneqq}\lim_{N\to\infty}\frac1N\,\mathbb{E}[\ln V_N]
\\
&=-\!\lim_{N\to\infty}\frac1N\,I(\bw^*;\bm\xi\, | \, X),
\end{aligned}
\label{eq:phi-MI}
\end{equation}
provided the limit exists.
Thus, $-\phi$ measures the information about the teacher supplied by
the labels per weight. Since each binary label conveys at most
$\ln2$ nats of information, this immediately yields
\begin{equation}
-\alpha\ln2\leq\phi\leq0.
\end{equation}

\section{Replica-symmetric solution}
We evaluate $\phi$ with the replica method, in which the pattern average reduces to an average over Gaussian local fields whose covariance is the replica overlap matrix, and we impose replica symmetry (RS) together with the Nishimori identity, which equates the teacher-student and student-student overlaps (Appendices~\ref{sec:replica} and~\ref{sec:nishimori}).
The {resulting RS prediction is}
\begin{equation}\label{eq:SaddlePointPhi}
{\phi_{\rm RS}(\alpha,\lambda)} =\max_{0\le r<1}\mathcal{F}_\lambda(r;\alpha)
\end{equation}
with
\begin{equation}
\mathcal{F}_\lambda(r;\alpha)=\frac{1}{2}\ln(1-r)+\frac{r}{2}-\alpha\,\E_{t\sim\mathcal{N}(0,1)}\,H_2(\psi_\lambda(r,t)).
\label{eq:F}
\end{equation}
Throughout the paper, $\ln$ denotes the natural logarithm; we will use
$\varphi(x)=\mathrm{e}^{-x^2/2}/\sqrt{2\pi}$ and
$\Phi(x)=\int_{-\infty}^{x}\varphi(u)\,\mathrm{d}u$
for the standard normal probability density and the cumulative
distribution function, respectively; we will use 
$\mathcal{D}t=\varphi(t)\,\mathrm{d}t$ for the standard
Gaussian integration measure. The function 
$H_2(p)=-p\ln p-(1-p)\ln(1-p)$ denotes the binary entropy
in nats, while $\mathcal{N}(\mu,v)$ will be reserved for a normal
distribution with mean $\mu$ and variance $v$.

The function $\psi_\lambda(r,t)$ appearing in~\eqref{eq:F} is defined by
\begin{equation}
\psi_\lambda(r,t)=\mathbb P_{z\sim\mathcal{N}(0,1)}\big[\sin\lambda\bigl(\sqrt r\,t+\sqrt{1-r}\,z\bigr)>0\big];
\label{eq:psi}
\end{equation}
it is the probability that a student with an overlap ${r=\bw\cdot\bw^*/N}$ with the teacher outputs $+1$, given the Gaussian mode $t$ it shares with the teacher. The maximizer $r_\star$ of \eqref{eq:SaddlePointPhi}, whenever it exists and is unique, is the RS prediction for the typical teacher-student overlap. The corresponding generalization error, defined as the error of a Gibbs student on a fresh input, is (see Appendix~\ref{sec:generr})
% \begin{align}
% \varepsilon_\lambda(r)&=2\,\E_{t\sim\mathcal{N}(0,1)}\big[\psi_\lambda(1-\psi_\lambda)\big]\nonumber\\
% &=\frac{1}{2}-\frac{8}{\pi^2}\!\!\sum_{a,b\ {\rm odd}}\!\frac{\sinh(\lambda^2 rab)}{ab}\,{\rm e}^{-\lambda^2(a^2+b^2)/2},
% \label{eq:eps}
% \end{align}
\begin{align}
\varepsilon_\lambda(r)
&=2\,\E_{t\sim\mathcal{N}(0,1)}\big[\psi_\lambda(1-\psi_\lambda)\big]
\nonumber\\
&=\frac{1}{2}-\frac{8}{\pi^2}
\sum_{m,n=0}^{\infty}
\frac{\sinh\bigl(\lambda^2 r(2m+1)(2n+1)\bigr)}
{(2m+1)(2n+1)}
\nonumber\\
&\qquad\qquad\qquad\quad \times
{\rm e}^{-\lambda^2((2m+1)^2+(2n+1)^2)/2},
\label{eq:eps}
\end{align}
which reduces to $\arccos(r)/\pi$ for $\lambda\to0^+$ and to $1/2$ for $\lambda\to\infty$ at fixed $r<1$, with exponentially small corrections that we bound explicitly in Appendix~\ref{sec:generr}. Equations~\eqref{eq:F} and~\eqref{eq:psi} coincide with the Bayes-optimal formula for generalized linear models with Gaussian prior and deterministic channel $y=s_\lambda(z)$~\cite{Barbier2019}, rigorously established there for Gaussian inputs and regular channels; for the present Rademacher inputs and discontinuous channel it rests on the standard replica assumptions detailed in Appendices~\ref{sec:replica} and~\ref{sec:nishimori}.

Because the Gaussian-smoothed square wave obeys a heat equation, the stationarity condition can be recast in the simple form (Appendix~\ref{sec:structure}),
\begin{align}
\partial_r\mathcal{F}_\lambda&=2\mathcal{J}_\lambda(r)\bigl(\alpha-\alpha_\lambda(r)\bigr),\nonumber\\
\alpha_\lambda(r)&=\frac{r}{4(1-r)\mathcal{J}_\lambda(r)},
\label{eq:stationary}
\end{align}
where 
\begin{equation}
\mathcal{J}_\lambda(r)=\frac{\lambda^2}{4}\E_{t\sim\mathcal{N}(0,1)}\bigg[\frac{(\partial_\theta\psi_\lambda)^2}{\psi_\lambda(1-\psi_\lambda)}\bigg]>0
\end{equation}
is $\lambda^2/4$ times the averaged Bernoulli Fisher information with respect to the mean phase $\theta=\lambda\sqrt r\,t$. 
At fixed $\lambda$ and training load $\alpha$, the stationary overlaps are the solutions of $\alpha_\lambda(r)=\alpha$. At each such point,
\[
\partial_r^2\mathcal{F}_\lambda(r;\alpha)
=-2\mathcal{J}_\lambda(r)\alpha_\lambda'(r).
\]
Since $\mathcal{J}_\lambda(r)>0$, solutions with
$\alpha_\lambda'(r)>0$ are local maxima of $\mathcal{F}_\lambda$,
whereas those with $\alpha_\lambda'(r)<0$ are local minima. 
The stationary curve $\alpha=\alpha_\lambda(r)$ also governs the fixed points  of Bayes-optimal generalized approximate message passing (GAMP)~\cite{Donoho2009,Rangan2011,Zdeborova2016}, an iterative inference algorithm that we use to numerically test the RS predictions for the stationary overlaps (Appendix~\ref{sec:GAMP}).
\section{Results}
\emph{Suppressed linear signal.}---For small loads, $r_\star=4\kappa_\lambda^2\,\alpha+O(\alpha^2)$ with
\begin{equation}
\kappa_\lambda=\tfrac12\E_z[z\,s_\lambda(z)]=\frac{2\lambda}{\pi}\sum_{k\ge0}{\rm e}^{-\lambda^2(2k+1)^2/2}.
\label{eq:kappa}
\end{equation}
The classical value $r_\star\simeq2\alpha/\pi$ is recovered for $\lambda\to0^+$, whereas $4\kappa_\lambda^2\simeq(16/\pi^2)\lambda^2{\rm e}^{-\lambda^2}$ for $\lambda\gg1$; each label is almost a fair coin for a student with $r\approx0$ and carries a nearly full bit, but only an exponentially small \emph{linear} signal about $\bw^*$. More generally, at any fixed $\alpha$ the RS maximizer satisfies $r_\star\to0$ and $\varepsilon_g\to1/2$ as $\lambda\to\infty$, {$-\phi_{\rm RS}\to\alpha\ln2$} (Appendix~\ref{sec:asymptotics}). The labels become maximally informative, yet a typical student learns nothing.

\begin{figure}[t]
\includegraphics[width=\columnwidth]{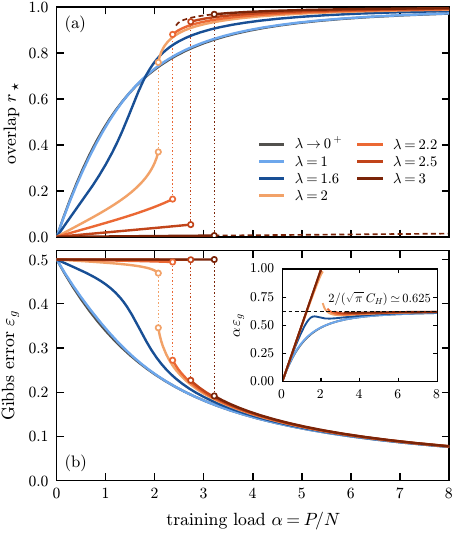}
\caption{Equilibrium RS learning curves. (a) Teacher-student overlap $r_\star$ and (b) Gibbs error $\varepsilon_g=\varepsilon_\lambda(r_\star)$ versus $\alpha$. For $\lambda>\lambda_c$ the curves jump at $\alpha_{\rm eq}(\lambda)$ (open circles: one-sided limits; dotted: jump). Thin dashed lines: metastable branches for $\lambda=3$. 
Inset: $\alpha\,\varepsilon_g$, approaching the $\lambda$-independent value $2/(\sqrt\pi C_H)\simeq0.625$ (dashed).}
\label{fig:curves}
\end{figure}

\begin{figure}[t]
\includegraphics[width=\columnwidth]{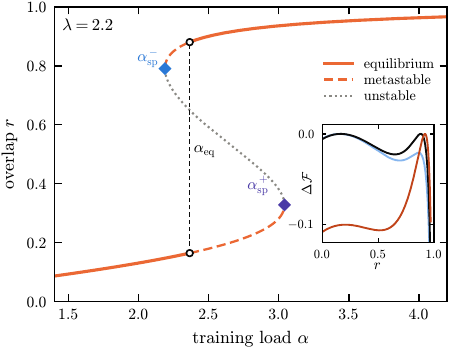}
\caption{Stationary points of $\mathcal{F}_\lambda$ at $\lambda=2.2$, i.e., the curve $\alpha=\alpha_\lambda(r)$ of Eq.~\eqref{eq:stationary}. Diamonds: spinodals; dashed vertical line: equilibrium jump at $\alpha_{\rm eq}=2.3639$. 
% Symbols: GAMP ($N=4000$, two instances per load) from uninformed (filled) and informed (open) initialization, displaying hysteresis. 
Inset: $\mathcal{F}_\lambda(r)-\max_r\mathcal{F}_\lambda$ at $\alpha=2.28$ (light blue), $\alpha_{\rm eq}$ (black), and $2.70$ (red).}
\label{fig:scurve}
\end{figure}

\emph{First-order learning transition.}---Figure~\ref{fig:curves} shows the equilibrium learning curves. For small $\lambda$, $\alpha_\lambda(r)$ is monotonic and $r_\star$ grows smoothly. At
\begin{equation}
(\lambda_c,\alpha_c,r_c)\simeq(1.94359,\,1.98100,\,0.5819),
\label{eq:critical}
\end{equation}
where $\alpha_\lambda'(r)=\alpha_\lambda''(r)=0$, the curve develops an inflection point, and for $\lambda>\lambda_c$ it becomes S-shaped (Fig.~\ref{fig:scurve}). Two folds, the spinodals $\alpha_\spin^-(\lambda)<\alpha_\spin^+(\lambda)$, delimit a load interval with two local maxima of $\mathcal{F}_\lambda$: A weakly aligned one, continuing the small-$\alpha$ branch, and a strongly aligned one. Their free entropies cross at the coexistence load $\alpha_{\rm eq}(\lambda)$, where the equilibrium overlap and error jump. For instance, at $\lambda=3$ one finds $\alpha_{\rm eq}=3.2125$, with $r_\star$ jumping from $0.0058$ to $0.9684$ and $\varepsilon_g$ from $0.49999$ to $0.1915$. We prove in Appendix~\ref{sec:structure} that, whenever two nondegenerate global maxima coexist, {$\phi_{\rm RS}$} is continuous and its load derivative jumps by
\begin{equation}
\Delta\partial_\alpha\phi_{\rm RS}=\E_{t\sim\mathcal{N}(0,1)}[H_2(\psi_\lambda(r_-,t))-H_2(\psi_\lambda(r_+,t))]>0.
\label{eq:jump}
\end{equation}
By the chain rule of mutual information,
$-\partial_\alpha\phi$ gives the predictive entropy of
a fresh label, whose RS prediction is
$-\partial_\alpha\phi_{\rm RS}$ at $\alpha_{\rm eq}$ the predictive uncertainty drops discontinuously, by $0.260$ nats at $\lambda=2.2$ and $0.387$ nats at $\lambda=3$. This is the analog, for continuous weights and a periodic readout, of the first-order transition to perfect generalization of the Ising perceptron~\cite{Gyorgyi1990,Sompolinsky1990}; in the classical spherical perceptron learning is always smooth~\cite{engel2001}. Selecting the global maximum is essential. Indeed, following a single stationary solution by continuation in $\alpha$ would retain a near-chance error well beyond $\alpha_{\rm eq}$.

\emph{Metastability and algorithms.}---Fig.~\ref{fig:scurve} shows the RS stationary branches. We notice that the observed metastability has an algorithmic counterpart. Indeed, in the interval $\alpha_{\mathrm{eq}}<\alpha<\alpha_{\mathrm{sp}}^{+}$, the strongly aligned solution is thermodynamically dominant, while GAMP from an uninformed initialization remains on the weakly aligned metastable branch. This gap between equilibrium and message-passing performance is characteristic of a “hard” phase in Bayesian inference~\cite{Zdeborova2016,RicciTersenghi2019}. The numerical GAMP comparison and its dependence on initialization are discussed in Appendix~\ref{sec:GAMP} (see Appendix~\ref{sec:numerics} for the numerical results).
\begin{figure}[t]
\includegraphics[width=\columnwidth]{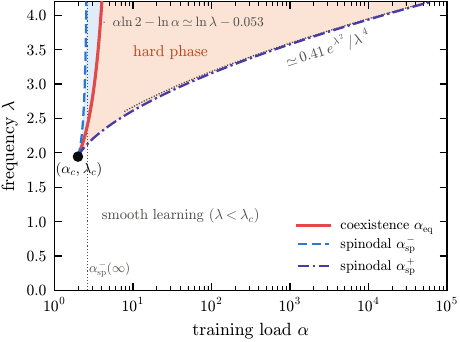}
\caption{RS phase diagram. Coexistence line $\alpha_{\rm eq}(\lambda)$ (solid), spinodals $\alpha_\spin^\pm(\lambda)$, and critical endpoint~\eqref{eq:critical}. Shading: aligned-phase metastability (blue) and hard phase $\alpha_{\rm eq}<\alpha<\alpha_\spin^+$ (orange). Dotted: large-$\lambda$ asymptotics~\eqref{eq:asym-sp} and~\eqref{eq:asym-eq}, and $\alpha_\spin^-(\infty)\simeq2.647$.}
\label{fig:phase}
\end{figure}

\emph{Scaling of the thresholds.}---The phase diagram in Fig.~\ref{fig:phase} reveals three different large-$\lambda$ behaviors, which we derive analytically in Appendix~\ref{sec:asymptotics}. On the weakly aligned branch, $r=O(\lambda^{-2})$ and only the first Fourier harmonic of $\psi_\lambda$ survives, giving $\mathcal{J}_\lambda(r)\simeq(4\lambda^2/\pi^2){\rm e}^{-\lambda^2}\cosh(\lambda^2r)$. The fold of $\alpha_\lambda(r)$ is then at $\lambda^2r=x_0$, with $x_0\tanh x_0=1$, and
\begin{equation}
\alpha_\spin^+\simeq\frac{\pi^2}{16}\,\frac{x_0}{\cosh x_0}\,\frac{{\rm e}^{\lambda^2}}{\lambda^4}\simeq0.409\,\frac{{\rm e}^{\lambda^2}}{\lambda^4},
\label{eq:asym-sp}
\end{equation}
with relative corrections $O(\lambda^{-2})$. On the aligned branch, each fringe $h=k\pi/\lambda$ acts as a classical perceptron boundary of width $\sqrt{(1-r)/r}$ in the variable $t$. Summing the boundary layers yields $\mathcal{F}_\lambda^{(+)}\simeq-\ln(\alpha\,C_H\Theta_\lambda)-\frac{1}{2}$, with $C_H=\int_{\mathbb R}H_2[\Phi(x)]\,\de x\simeq1.8064$ ($\Phi$ is the cumulative distribution function of the standard normal distribution) and $\Theta_\lambda=\sum_{k\in\mathbb Z}\varphi(k\pi/\lambda)\to\lambda/\pi$ ($\varphi$ the normal density). Equating it with the weakly aligned value $\mathcal{F}_\lambda^{(-)}\simeq-\alpha\ln2$ gives
\begin{equation}
\alpha_{\rm eq}\ln2-\ln\alpha_{\rm eq}\simeq\ln\lambda+\ln\frac{C_H}{\pi}+\frac{1}{2},
\label{eq:asym-eq}
\end{equation}
i.e., $\alpha_{\rm eq}\simeq\log_2\lambda+\log_2\log_2\lambda$, which reproduces the exact coexistence line within $1.3\%$ at $\lambda=3$ and $0.2\%$ at $\lambda=5$. Finally, the aligned branch appears at $\alpha_\spin^-\to2.647$. The logarithmic law has a simple information-theoretic reading: To generalize, the student must localize the teacher within one fringe, which costs $\simeq\ln\lambda$ nats per weight, while each label provides at most $\ln 2$. Statistically, a high frequency is thus cheap. Algorithmically it is not, i.e.,  the linear signal~\eqref{eq:kappa} that guides an uninformed learner is $\propto {\rm e}^{-\lambda^2}$, so the hard phase widens super-exponentially in $\lambda$. This mirrors, in a proportional high-dimensional limit and at fixed frequency, the cryptographic hardness of learning periodic neurons whose frequency grows with dimension~\cite{Song2021,Bruna2021}, and it complements the overlap-gap analysis of the square-wave perceptron with binary weights and Gaussian inputs~\cite{Benedetti2025,Gamarnik2021}.

\emph{Universal asymptotics and fragmented version space.}---Once the aligned phase is reached, the boundary-layer solution gives $\varepsilon_g\simeq2/(\sqrt\pi\,C_H\,\alpha)\simeq0.625/\alpha$ for every $\lambda$ [inset of Fig.~\ref{fig:curves}(b)], the classical Gibbs-learning law~\cite{Seung1992,engel2001}. The frequency enters only through $\Theta_\lambda$, which rescales $1-r_\star$ and shifts the mutual information by $\ln[\sqrt{2\pi}\,\Theta_\lambda]$, but leaves the error unchanged. Hence, more fringes make each label more informative about $\bw^*$ and, in the same proportion, more fragile to misalignment. The version space also has a markedly nonclassical geometry. Since a field can move between two fringes of equal label only by crossing a fringe of opposite label, $\mathcal{V}$ is the disjoint union of cells labeled by the fringe index $k_\mu$ of every field, whereas for $\lambda\to0^+$ it is convex. Any continuous zero-training-error dynamics is confined to its initial cell; accordingly, samplers started at the teacher remain in the teacher's cell and overestimate $r_\star$ (Appendix~\ref{sec:fragmentation}).

\section{Discussions}
The periodic measurement response of a quantum perceptron has opposite consequences for storage and for learning. It increases the number of random patterns that can be stored~\cite{BGGMP26}, but it turns generalization into a first-order transition, delays its onset logarithmically in the frequency, and makes it algorithmically hard over a range of loads that grows as ${\rm e}^{\lambda^2}/\lambda^4$, while leaving the asymptotic error unchanged. Since the output of any parametrized circuit is a trigonometric polynomial whose frequencies are fixed by the architecture~\cite{SchuldSwekeMeyer2021}, and data re-uploading increases them~\cite{PerezSalinas2020}, we expect analogous transitions whenever high-frequency components dominate a quantum model's decision function. Therefore, the spectrum that boosts expressivity also sets the sample and computational complexity of learning. Open problems include a proof of the RS formula for binary inputs and discontinuous channels, the effect of finite-shot readout and of prior or model mismatch, and the analysis of training algorithms on the fragmented version space.

\section*{Acknowledgments}
G.G. is partially supported by INFN through the project ``QUANTUM'', by the Italian National Group of Mathematical Physics (GNFM-INdAM), and by the University of Bari through the 2023-UNBACLE-0245516 grant. M.G. acknowledges financial support from the European Union under ERC Advanced Grant TAtypic, Project No.~101142236. Views and opinions expressed are those of the authors only and do not necessarily reflect those of the European Union or the European Research Council Executive Agency.

\clearpage
\appendix
\setcounter{secnumdepth}{2}
\onecolumngrid

%=====================================================================
\section{Circuit and classical reduction}\label{sec:circuit}
%=====================================================================
Here we derive the sinusoidal expectation value of the readout observable in~\eqref{eq:readout}, and show how its sign yields the periodic classifier defined in~\eqref{eq_sin}. We also establish the absence of input–output entanglement for basis-encoded inputs, and explain the model’s classical reduction under the assumption of exact expectation values.

We denote by $|\pm1\rangle$ the eigenstates of $\sigma_z$
with eigenvalues $\pm1$, respectively. The input $\bx\in\{\pm1\}^N$ {is basis encoded into
$|\bx\rangle=|x_1\rangle\otimes\cdots\otimes|x_N\rangle$,
where} $\sigma_z|x_j\rangle=x_j|x_j\rangle$; the readout qubit is prepared in the $\sigma_z$ eigenstate $|+1\rangle$ with eigenvalue $+1$. The trainable gate is
\begin{equation}
U_\lambda(\bw)=\exp\Bigl(-\frac{\ii}{2}\lambda\,\hat\bw\cdot\bm\sigma_z\otimes\sigma_y^{(\rm out)}\Bigr),\qquad \hat\bw\cdot\bm\sigma_z=\sum_{j=1}^N\frac{w_j}{\|\bw\|}\sigma_z^{(j)},
\label{eq:S-U}
\end{equation}
with $\lambda>0$ a fixed architectural parameter. Since $|\bx\rangle$ is a joint eigenvector of the $\sigma_z^{(j)}$, the gate rotates the readout by $\theta=\lambda\,\bw\cdot\bx/\|\bw\|$, giving
\begin{equation}
U_\lambda(\bw)\,|\bx\rangle\otimes|+1\rangle=|\bx\rangle\otimes {\rm e}^{-\ii\theta\sigma_y^{(\mathrm{out})}/2}|+1\rangle
=|\bx\rangle\otimes\bigl(\cos(\theta/2)|+1\rangle
+\sin(\theta/2)|-1\rangle\bigr).
\end{equation}
The expectation value of $\sigma_x^{(\rm out)}$ is therefore
$2\cos(\theta/2)\sin(\theta/2)=\sin\theta$, as in
Eq.~\eqref{eq:readout}. The unitary depends only on the direction of $\bw$; we fix the redundant radial degree of freedom by $\|\bw\|=\sqrt N$, so that $\theta=\lambda h$ with $h=\bw\cdot\bx/\sqrt N=O(1)$. The binary classifier is
\begin{equation}
g_\bw(\bx)=s_\lambda\Bigl(\frac{\bw\cdot\bx}{\sqrt N}\Bigr),\qquad s_\lambda(h)=\sgn\bigl(\sin(\lambda h)\bigr),
\label{eq:S-classifier}
\end{equation}
whose value on the measure-zero set $\sin(\lambda h)=0$ is irrelevant below.

Three features of generic variational classifiers are retained: classical data are mapped to quantum states, trainable parameters enter a unitary, and the prediction is a thresholded single-qubit expectation value~\cite{FarhiNeven2018,Mitarai2018}. The Fourier structure of parametrized expectation values~\cite{SchuldSwekeMeyer2021} collapses here to the single mode $\sin(\lambda\hat\bw\cdot\bx)$, whose sign changes on the parallel hyperplanes $\hat\bw\cdot\bx=k\pi/\lambda$, $k\in\Z$. Conversely, for basis inputs the circuit generates no input-output entanglement, all input generators commute, and exact expectation values (no shot noise) are assumed. The effects discussed in the main text are therefore properties of the periodic measurement response, not of entanglement or of a quantum computational advantage.

%=====================================================================
\section{Teacher-student setting, evidence and generalization}\label{sec:setting}
%=====================================================================

{We formulate here the learning process in the teacher–student framework \cite{engel2001}, focusing on the “generalization” (probability of error made by student on fresh inputs).}

The weights are drawn from the uniform measure on $\mathbb S^{N-1}(\sqrt N):=\{\bw\in \R^N: \|\bw\|=\sqrt{N}\}$,
\begin{equation}
\de\mu(\bw)=\frac{1}{Z_N}\,\de^N\bw\,\delta\bigl(\|\bw\|^2-N\bigr),\qquad Z_N=\frac{\pi^{N/2}N^{N/2-1}}{\Gamma(N/2)}.
\label{eq:S-prior}
\end{equation}
A teacher $\bw^*\sim\mu$ is drawn once; the inputs $X=(\bx^1,\dots,\bx^P)$ have i.i.d.\ Rademacher entries, i.e. $\PP(x_i^\mu=\pm1)=1/2$; the labels are generated according to the teacher as $\xi^\mu=g_{\bw^*}(\bx^\mu)$ and we denote with $\mathcal{T}_P=\{(\bx^\mu,\xi^\mu)\}_{\mu=1}^P$ the corresponding training set. We consider the thermodynamic limit $N,P\to\infty$ with $\alpha=P/N$ fixed. The version space $\mathcal{V}(\mathcal{T}_P)=\{\bw:\,g_\bw(\bx^\mu)=\xi^\mu\ \text{for all }\mu\}$ of weights compatible with the training set has (Gardner) volume~\cite{gard1987,gard1988,gard1988s}
\begin{equation}
V_N(\mathcal{T}_P)=\int\de\mu(\bw)\prod_{\mu=1}^P\mathbf 1\{g_\bw(\bx^\mu)=\xi^\mu\},
\label{eq:S-volume}
\end{equation}
which is strictly positive almost surely, since a neighborhood of the teacher is compatible. The uniform measure over the set of weights compatible with the training set is given by
\begin{equation}
\de\mu_{\mathcal{T}_P}(\bw)=\frac{\de\mu(\bw)}{V_N(\mathcal{T}_P)}\prod_{\mu=1}^P\mathbf 1\{g_\bw(\bx^\mu)=\xi^\mu\}.
\label{eq:S-posterior}
\end{equation}
The quenched entropy density is
\begin{equation}
\phi(\alpha,\lambda)=\lim_{N\to\infty}\frac1N\E\ln V_N
\label{eq:S-phi}
\end{equation}
(assuming the limit exists).

For fixed weights, let $\mathcal{R}_N(\bw,\bw^*)=\mathbb P_\bx[g_\bw(\bx)\neq g_{\bw^*}(\bx)]$ and $r_N=\bw\cdot\bw^*/N$. If $r_N\to r$ and both vectors are delocalized, $\max_j|w_j|/\sqrt N\to0$ and $\max_j|w^*_j|/\sqrt N\to0$, the joint central limit theorem (see section~\ref{sec:replica}, equation~\eqref{eq:S-Zn}) gives $\mathcal{R}_N\to\varepsilon_\lambda(r)$, with
\begin{equation}
\varepsilon_\lambda(r)=\mathbb P\big[s_\lambda(U)\ne s_\lambda(V)\big],\qquad (U,V)\ \text{centered Gaussian},\ \E U^2=\E V^2=1,\ \E UV=r.
\label{eq:S-epsdef}
\end{equation}
Delocalization holds with probability tending to one under $\mu$, and also for samples from $\mu_{\mathcal T_P}$ after averaging over the planted data (see subsection~\ref{sec:nishimori}). The risk of a student sampled from $\mu_{\mathcal T_P}$ is therefore
\begin{equation}
\varepsilon_{g,N}=\int\de\mu_{\mathcal{T}_P}(\bw)\,\mathcal{R}_N(\bw,\bw^*)=\int_{-1}^1\varepsilon_\lambda(r)\,\de\nu_{\mathcal{T}_P,\bw^*}(r)+o(1),
\label{eq:S-gibbsrisk}
\end{equation}
where $\nu_{\mathcal{T}_P,\bw^*}$ is the push-forward of $\mu_{\mathcal T_P}$ under $\bw\mapsto\bw\cdot\bw^*/N$. If this overlap measure concentrates on $r_\star(\alpha,\lambda)$, then $\varepsilon_g=\varepsilon_\lambda(r_\star)$. At a coexistence point the limiting overlap measure may have two atoms whose weights are not fixed by the leading-order free entropy; we then report both one-sided limits.

%=====================================================================
\subsection{Replica calculation}\label{sec:replica}
%=====================================================================

{Here, we evaluate the expectation of the Gardner volume in the thermodynamic limit with the replica method, in which the pattern average reduces to an average over Gaussian local fields whose covariance is the replica overlap matrix. We impose replica symmetry (RS).}

We use the replica method, consisting of starting from the identity
\begin{equation}
    \E\ln V_N=\lim_{n\to0}\frac{1}{n}\ln\E V_N^n,
\end{equation}
compute $\E V_N^n$ for integer $n$, take $N\to\infty$ at fixed $\alpha$, and use the result obtained as an analytic continuation to $n\to0$. Regarding the teacher as replica $0$, $\bW^0=\bw^*$, $\bW^a=\bw^a$ ($a=1,\dots,n$), independence over patterns gives
\begin{equation}
\E V_N^n=\int\prod_{a=0}^n\de\mu(\bW^a)\,\big[\mathcal{A}_{N,\lambda}(\bW)\big]^P,
\end{equation}
where
\begin{equation}
\mathcal{A}_{N,\lambda}(\bW)=\E_\bx\prod_{a=1}^n\Theta\bigl(\sin(\lambda h_0)\sin(\lambda h_a)\bigr),\quad h_a\equiv h(\bm{W}^a,\bm{x})=\frac{\bW^a\cdot\bx}{\sqrt N}.
\label{eq:S-replicated}
\end{equation}
The $N$ components of the pattern $\bm{x}$ enter into $\mathcal{A}_{N,\lambda}(\bm{W})$ only through the
\(n+1\) ``local fields'' $\bm{h}=(h_0,h_1,\dots,h_n)$. For fixed weight vectors, the random pattern induces a joint distribution of
the field vector which can be formally written in the form
\begin{equation}\label{eq:induced-field-distribution}
\rho_N(\bm h| \bm{W})\coloneqq\E_{\bm x}\prod_{a=0}^{n}
\delta\! \left(h_a-\frac{\bW^a\cdot\bx}{\sqrt N}\right),
\end{equation}
with equality to be understood in the distributional sense. The single-pattern contribution $\mathcal{A}_{N,\lambda}(\bm{W})$ in equation~\eqref{eq:S-replicated} can then be rewritten exactly as
\begin{equation}\label{eq:one-pattern-field-average}
\mathcal{A}_{N,\lambda}(\bm{W}) = \int_{\mathbbm{R}^{n+1}}\dd \bm{h}\,\rho_N(\bm h|\bm{W})\prod_{a=1}^{n}
\Theta\big(\sin(\lambda h_0)\sin(\lambda h_a)\big).
\end{equation}
A standard computation (see Section~\ref{sec:ConvergenceLocalFields}) shows that in the high-dimensional limit $N\rightarrow \infty$, $\rho_N(\bm{h}|\bm{W})$ converges to a centered Gaussian distribution with a correlation matrix given by the overlap (or Gram) matrix
\begin{equation}
  Q_{ab}\coloneqq \frac{\bm W^a\cdot\bm W^b}{N},
  \qquad a,b=0,1,\dots,n.
  \label{eq:Q-definition}
\end{equation}
% Fix $n$ and a sequence of weights with overlap matrix $Q_{ab}=\bW^a\cdot\bW^b/N$ ($Q_{aa}=1$) converging to $Q$ and with $\max_j|W^a_j|/\sqrt N\to0$ for all $a$. The characteristic function of $H=(H_0,\dots,H_n)$ is
% \begin{equation}
% \varphi_N(\mathbf k)=\E_\bx {\rm e}^{\ii\mathbf k\cdot\bh}=\prod_{j=1}^N\cos b_{N,j},\qquad b_{N,j}=\frac{1}{\sqrt N}\sum_{a=0}^nk_aW^a_j.
% \end{equation}
% Using $\ln\cos y=-y^2/2+O(y^4)$ and $\sum_jb_{N,j}^4\le(\max_jb_{N,j}^2)\sum_jb_{N,j}^2\to0$, one finds $\ln\varphi_N(\mathbf k)\to-\frac{1}{2}\mathbf k^{\mathsf T}Q\mathbf k$, and L\'evy's continuity theorem gives weak convergence $\bh\to\mathcal{N}(\mathbf 0,Q)$. The discontinuities of the integrand in Eq.~\eqref{eq:S-replicated} lie on the hyperplanes $h_a=k\pi/\lambda$, which have zero probability under the limiting law (every field has unit variance), hence
Therefore
\begin{equation}
\mathcal{A}_{N,\lambda}(\bW)\to\mathcal{Z}_{n,\lambda}(Q)=\int\de\gamma_Q(\bh)\prod_{a=1}^n\Theta\bigl(\sin(\lambda h_0)\sin(\lambda h_a)\bigr),
\label{eq:S-Zn}
\end{equation}
with
\begin{equation}\label{eq:GaussianMeasureCorrelated}
{\de\gamma_Q(\bh)\coloneqq\frac{\exp(-\frac{1}{2}\bh^T Q^{-1} \bh)}{(2\pi)^{(n+1)/2}(\det Q)^{1/2}}\de \bh.}
\end{equation}
The resulting integrand depends on the weights only through their
overlaps. We can therefore organize the integration by grouping all
configurations that realize the same matrix $Q$. Their normalized
volume is described by the geometric density
\begin{equation*}
\mathcal J_N(Q)\coloneqq
\int\prod_{a=0}^n\de\mu(\bW^a)
\prod_{0\leq a<b\leq n}
\delta\!\left(Q_{ab}-\frac{\bW^a\cdot\bW^b}{N}\right).
\end{equation*}
This is the probability density of the overlap matrix under the
product spherical measure. Only off-diagonal constraints appear in the product of delta
functions, since the spherical measures already impose $Q_{aa}=1$.
Consequently,
\begin{equation*}
\E V_N^n\asympeq
\int_{\mathcal Q}\de Q\,\mathcal J_N(Q)
\bigl[\mathcal Z_{n,\lambda}(Q)\bigr]^{\alpha N}.
\end{equation*}

The standard Gram-matrix integration formula~\cite[Appendix~B]{Fyodorov2010}
gives, for $Q>0$,
\begin{equation*}
\mathcal J_N(Q)=C_{N,n}\,\bigl(\det Q\bigr)^{(N-n-2)/2},
\end{equation*}
where $C_{N,n}$ is independent of $Q$ and
$\ln C_{N,n}=O(\ln N)$ at fixed $n$. Thus at the leading contribution in the exponential rate: 
\begin{equation}
\frac{1}{N}\ln\!\left(
\mathcal J_N(Q)\bigl[\mathcal Z_{n,\lambda}(Q)\bigr]^{\alpha N}
\right)
={}\frac{1}{2}\ln\det Q
+\alpha\ln\mathcal Z_{n,\lambda}(Q)+o(1),
\end{equation}
for each fixed positive-definite $Q$. Then a saddle-point approximation selects
the largest exponential rate over the admissible overlap matrices,
giving formally
\begin{equation}
\lim_{N\to\infty}\frac1N\ln\E V_N^n=\sup_{Q\geq 0,\,Q_{aa}=1}\Bigl(\frac{1}{2}\ln\det Q+\alpha\ln\mathcal{Z}_{n,\lambda}(Q)\Bigr).
\label{eq:S-saddle}
\end{equation}

\subsection{{Replica-Symmetric ansatz}}\label{sec:RS-ansatz}
{In this section, we specialize the search of the saddle point \eqref{eq:S-saddle} to the replica symmetric subspace, that is, we put all the student-student overlaps equal to $q$ and all the teacher-student overlaps equal to $r$:
$(Q_{\rm RS})_{0a}=r$ and $(Q_{\rm RS})_{ab}=q$ for $1\le a<b\le n$. Using a Schur complement decomposition, the determinant is given by}
\begin{equation}
\det Q_{\rm RS}=(1-q)^{n-1}\bigl(1+(n-1)q-nr^2\bigr)=(1-q)^n\Bigl(1+n\frac{q-r^2}{1-q}\Bigr),
\end{equation}
{and, in the limit $n\rightarrow 0$:
\begin{equation}
\frac{1}{2n}\ln\det Q_{\rm RS}=\frac{1}{2}\ln(1-q)+\frac{q-r^2}{2(1-q)}+O(n).
\label{eq:S-det}
\end{equation}}
For the pattern term, the identity $\Theta(\sin\lambda h_0\sin\lambda h_a)=\Theta(\sin\lambda h_0)\Theta(\sin\lambda h_a)+\Theta(-\sin\lambda h_0)\Theta(-\sin\lambda h_a)$ and the invariance of $\gamma_Q$ under $\bh\mapsto-\bh$ give 
\begin{equation}
\mathcal{Z}_{n,\lambda}=2\int\de\gamma_Q\,\Theta(\sin\lambda h_0)\prod_{a\ge1}\Theta(\sin\lambda h_a).
\end{equation}
For $0<q<1$ and $r^2\leq q$, the Gaussian fields $\bh$ with
correlation matrix $Q_{\rm RS}$ admit the representation
\begin{equation}
h_0=\frac{r}{\sqrt q}\,t+\sqrt{1-\frac{r^2}{q}}\,y,\qquad h_a=\sqrt q\,t+\sqrt{1-q}\,z_a,\qquad t,y,z_a\ \text{i.i.d.}\ \mathcal{N}(0,1),
\label{eq:S-rep}
\end{equation}
whose explicit derivation is given in
Appendix~\ref{sec:GaussianDecomposition}. There we decompose the
RS covariance matrix, remove the residual coupling between replicas
with an auxiliary Gaussian variable, and rotate the two shared modes
to obtain the single common variable $t$. Conditionally on $t$, all
fields are independent, and
\begin{equation}
\mathcal{Z}_{n,\lambda}(q,r)=2\int\Dt\,\chi_\lambda(q,r,t)\,\psi_\lambda(q,t)^n,\qquad
\psi_\lambda(q,t)=p_\lambda(\sqrt q\,t,1-q),\quad \chi_\lambda(q,r,t)=p_\lambda\Bigl(\frac{rt}{\sqrt q},1-\frac{r^2}{q}\Bigr),
\label{eq:S-ZRS}
\end{equation}
in terms of the Gaussian-smoothed square wave. Its probabilistic
definition applies for $v\geq0$, while the series below requires
$v>0$:
\begin{equation}
p_\lambda(\mu,v)=\E_{z\sim\mathcal{N}(0,1)}\,\Theta\big[\sin\lambda(\mu+\sqrt v\,z)\big]
=\frac{1}{2}+\frac2\pi\sum_{k=0}^\infty\frac{\sin\bigl((2k+1)\lambda\mu\bigr)}{2k+1}\,{\rm e}^{-(2k+1)^2\lambda^2v/2}\qquad(v>0).
\label{eq:S-p}
\end{equation}
At $q=r^2>0$, the teacher probability is evaluated with
$p_\lambda(\mu,0)=\Theta(\sin\lambda\mu)$. The case $q=r=0$ is
evaluated directly, since all fields are then independent and
$\mathcal Z_{n,\lambda}(0,0)=2^{-n}$.
The series follows from the Fourier expansion of $\Theta(\sin x)$ and $\E\sin(a(\mu+\sqrt v z))=\sin(a\mu){\rm e}^{-a^2v/2}$; it converges absolutely and uniformly in $\mu$ for $v>0$. Using $\int\Dt\,\chi_\lambda=1/2$ (from $p_\lambda(-\mu,v)=1-p_\lambda(\mu,v)$) and $\psi^n=1+n\ln\psi+O(n^2)$,
\begin{equation}
\phi_{\rm RS}(\alpha,\lambda)=\operatorname*{extr}_{0\le r^2\le q<1}\mathcal{K}_\lambda(q,r;\alpha),\qquad
\mathcal{K}_\lambda(q,r;\alpha)=\frac{1}{2}\ln(1-q)+\frac{q-r^2}{2(1-q)}+2\alpha\int\Dt\,\chi_\lambda(q,r,t)\ln\psi_\lambda(q,t).
\label{eq:S-K}
\end{equation}
\subsection{Bayesian interpretation, Nishimori identity and information}
\label{sec:nishimori}
%=====================================================================

The measure $\mu_{\mathcal T_P}$ defined in Eq.~\eqref{eq:S-posterior}
was introduced as the uniform measure on the version space. We now
explain its Bayesian interpretation and derive the identities used
in the RS reduction and in the information-theoretic interpretation
of the learning curves. The Bayesian identities below are exact at
finite $N$; the additional assumptions needed to pass to the RS
description will be stated separately.

\paragraph{Likelihood, evidence and posterior.}
The spherical measure $\mu$ is the \emph{prior}: it describes the
teacher weights before any labels are observed. The inputs $X$ are
independent of the teacher. To distinguish a possible label sequence
from the random labels $\bm\xi=(\xi^1,\ldots,\xi^P)$, let
$\bm\eta\in\{\pm1\}^P$ and write
\begin{equation}
\mathcal T_P(X,\bm\eta)
\coloneqq\{(\bx^\mu,\eta^\mu)\}_{\mu=1}^P.
\end{equation}
Thus the observed training set is
$\mathcal T_P=\mathcal T_P(X,\bm\xi)$. Given a candidate teacher
$\bw$, the probability of the label sequence $\bm\eta$, called its
\emph{likelihood}, is
\begin{equation}
\PP(\bm\xi=\bm\eta\mid X,\bw^*=\bw)
=\prod_{\mu=1}^P\one\{g_\bw(\bx^\mu)=\eta^\mu\}.
\label{eq:S-bayes-likelihood}
\end{equation}
It equals one when all labels are reproduced and zero otherwise,
because the labeling rule is deterministic. Integrating over the
teacher prior gives the probability of those labels at fixed inputs:
\begin{equation}
\PP(\bm\xi=\bm\eta\mid X)
=\int\de\mu(\bw)\prod_{\mu=1}^P
\one\{g_\bw(\bx^\mu)=\eta^\mu\}
=V_N\bigl(\mathcal T_P(X,\bm\eta)\bigr).
\label{eq:S-bayes-evidence}
\end{equation}
This marginal probability is the \emph{evidence}. The same Gardner
volume therefore measures both the fraction of compatible weights
and the probability assigned to the observed label sequence under the uniform prior on the teacher weights.

Bayes' rule gives the conditional distribution of the teacher after
observing $\mathcal T_P$. For every measurable subset $A$ of the
weight sphere,
\begin{equation}
\PP(\bw^*\in A\mid\mathcal T_P)
=\frac{1}{V_N(\mathcal T_P)}\int_A\de\mu(\bw)
\prod_{\mu=1}^P\one\{g_\bw(\bx^\mu)=\xi^\mu\}
=\mu_{\mathcal T_P}(A).
\label{eq:S-teacher-posterior}
\end{equation}
Thus the \emph{posterior} is exactly
the already defined measure $\mu_{\mathcal T_P}$, the measure defining the version space, and a student sampled from this measure is
what has been called a \emph{Gibbs student}. In this inference model the student then uses the same prior and likelihood that generated
the data, and this matched setting is called \emph{Bayes-optimal
inference}. The Nishimori identity is a consequence of this matched setting~\cite{nish2001,Iba1999,Zdeborova2016}, and its relevant content here is the equality of conditional distributions in
\eqref{eq:S-teacher-posterior}. We illustrate the consequences of this identity in the following.

For a fixed training set, let $\bw^1,\ldots,\bw^k$ be independent
samples from $\mu_{\mathcal T_P}$. We denote their conditional
average by brackets:
\begin{equation}
\big\langle\mathcal O(\bw^1,\ldots,\bw^k)\big\rangle_{\mathcal T_P}
\coloneqq
\int\prod_{a=1}^k\de\mu_{\mathcal T_P}(\bw^a)\,
\mathcal O(\bw^1,\ldots,\bw^k).
\label{eq:S-posterior-average}
\end{equation}
If the observable also contains $\bw^*$, the brackets integrate only
the student variables, keeping the teacher fixed. A further
expectation can be taken on this quantity, which we denote by $\E$, when we want to take averages over the teacher and the training inputs,
with the labels generated by the teacher. For quantities depending
only on $\mathcal T_P$, this is equivalently the average
$\E_{\mathcal T_P}$ over training sets. 

By Eq.~\eqref{eq:S-teacher-posterior}, conditioned on
$\mathcal T_P$ the teacher also has law $\mu_{\mathcal T_P}$.
It is conditionally independent of the 
students sampled from $\mu_{\mathcal{T}_P}$, so their joint conditional law is the product of $k+1$
copies of $\mu_{\mathcal T_P}$. The teacher can therefore be exchanged
with an additional posterior sample. For every bounded observable
$\mathcal O$, this gives the Nishimori identity
\begin{equation}
\E\big\langle
\mathcal O(\bw^*,\bw^1,\ldots,\bw^k;\mathcal T_P)
\big\rangle_{\mathcal T_P}
=\E_{\mathcal T_P}\big\langle
\mathcal O(\bw^{k+1},\bw^1,\ldots,\bw^k;\mathcal T_P)
\big\rangle_{\mathcal T_P}.
\label{eq:S-nishimori}
\end{equation}
Indeed, conditioning the left-hand side on $\mathcal T_P$ integrates
the teacher against $\mu_{\mathcal T_P}$ as well; renaming that
integration variable $\bw^{k+1}$ gives the right-hand side.

Choosing any bounded function $f$ of the overlap gives
\begin{equation}
\E\left\langle
f\!\left(\frac{\bw^*\cdot\bw^1}{N}\right)
\right\rangle_{\mathcal T_P}
=\E\left\langle
f\!\left(\frac{\bw^2\cdot\bw^1}{N}\right)
\right\rangle_{\mathcal T_P}.
\label{eq:S-overlap-laws}
\end{equation}
Hence the teacher--student and student--student overlaps have the
same disorder-averaged distribution. However, this does not assert equality
of the overlaps for individual weight vectors. Moreover, conditional
independence of two posterior samples gives
\begin{equation}
\frac1N\E\bigl[\bw^*\cdot\langle\bw\rangle_{\mathcal T_P}\bigr]
=\frac1N\E\langle\bw^1\cdot\bw^2\rangle_{\mathcal T_P}
=\frac1N\E\bigl\|\langle\bw\rangle_{\mathcal T_P}\bigr\|^2
\geq0,
\label{eq:S-mean-overlap}
\end{equation}
where $\langle\bw\rangle_{\mathcal T_P}
=\int\de\mu_{\mathcal T_P}(\bw)\,\bw$ is the posterior mean.
If the two overlaps concentrate on deterministic values $r$ and $q$,
as assumed in the RS description, Eqs.~\eqref{eq:S-overlap-laws}
and~\eqref{eq:S-mean-overlap} imply $r=q\geq0$.
However, the identity itself does not establish overlap concentration,
uniqueness of an equilibrium overlap, or stability against
replica-symmetry breaking. Equation~\eqref{eq:S-mean-overlap} also
explains the relation between the averaged posterior-mean and
posterior-sample overlaps used in Appendix~\ref{sec:GAMP}. Since
$\|\bw^*\|^2=N$, it gives
\begin{equation}
\frac1N\E\bigl\|\bw^*-\langle\bw\rangle_{\mathcal T_P}\bigr\|^2
=1-\frac1N\E\bigl\|\langle\bw\rangle_{\mathcal T_P}\bigr\|^2.
\end{equation}

A second consequence, obtained by averaging
Eq.~\eqref{eq:S-teacher-posterior}, is
\begin{equation}
\E_{\mathcal T_P}\mu_{\mathcal T_P}(A)=\mu(A)
\label{eq:S-posterior-marginal}
\end{equation}
for every measurable set $A$ independent of the data. Thus, after
averaging over training sets, a posterior sample has the original
spherical law. In particular, the delocalization property of the
spherical weights also holds for posterior samples in the planted
ensemble. Together with the joint Gaussian limit in
Appendix~\ref{sec:ConvergenceLocalFields}, this justifies the use of
$\varepsilon_\lambda(r)$ in Eq.~\eqref{eq:S-gibbsrisk}; its remainder
vanishes in probability and in mean absolute value over the teacher
and training data. The conditional exchangeability also explains
why the teacher is a typical posterior sample, as used in
Appendix~\ref{sec:fragmentation}.

\paragraph{Reduction of the RS potential.}
We now apply $q=r$ to the RS functional
$\mathcal K_\lambda(q,r;\alpha)$ of Eq.~\eqref{eq:S-K}. For
$0<q<1$, the definitions in Eq.~\eqref{eq:S-ZRS} give
\begin{equation}
\chi_\lambda(q,q,t)=\psi_\lambda(q,t),
\qquad
\left.\partial_r\chi_\lambda(q,r,t)\right|_{r=q}
=2\partial_q\psi_\lambda(q,t),
\qquad
\left.\partial_q\chi_\lambda(q,r,t)\right|_{r=q}
=-\partial_q\psi_\lambda(q,t).
\label{eq:S-diagonal-derivatives}
\end{equation}
The partial derivatives are taken before setting $r=q$. For example,
\begin{equation}
\partial_q\psi_\lambda(q,t)
=\frac{t}{2\sqrt q}\,
\partial_\mu p_\lambda(\sqrt q\,t,1-q)
-\partial_v p_\lambda(\sqrt q\,t,1-q),
\end{equation}
and differentiating the two arguments of $\chi_\lambda$ gives
Eq.~\eqref{eq:S-diagonal-derivatives}. The symmetry
$\psi_\lambda(q,-t)=1-\psi_\lambda(q,t)$ implies
\begin{equation}
\int\mathcal D t\,\psi_\lambda(q,t)=\frac12,
\qquad
\int\mathcal D t\,\partial_q\psi_\lambda(q,t)=0.
\end{equation}
Consequently,
\begin{equation}
\begin{aligned}
\left.\partial_r\mathcal K_\lambda(q,r;\alpha)\right|_{r=q}
&=-\frac{q}{1-q}
+4\alpha\int\mathcal D t\,
\partial_q\psi_\lambda(q,t)\ln\psi_\lambda(q,t),\\
\left.\partial_q\mathcal K_\lambda(q,r;\alpha)\right|_{r=q}
&=-\frac12
\left.\partial_r\mathcal K_\lambda(q,r;\alpha)\right|_{r=q}.
\end{aligned}
\label{eq:S-diagonal-stationarity}
\end{equation}
Thus both stationarity equations reduce to the same condition.
Writing the common overlap as $r$, it reads
\begin{equation}
\frac{r}{1-r}
=4\alpha\int\mathcal D t\,
\partial_r\psi_\lambda(r,t)\ln\psi_\lambda(r,t).
\label{eq:S-saddle-reduced}
\end{equation}
This is also the stationarity condition of
$\mathcal F_\lambda(r;\alpha)=\mathcal K_\lambda(r,r;\alpha)$,
because its derivative is the sum of the two partial derivatives in
Eq.~\eqref{eq:S-diagonal-stationarity}. Explicitly,
\begin{equation}
\begin{aligned}
\mathcal F_\lambda(r;\alpha)
&=\frac12\ln(1-r)+\frac r2+2\alpha\mathcal I_\lambda(r),\\
\mathcal I_\lambda(r)
&=\int\mathcal D t\,\psi_\lambda(r,t)\ln\psi_\lambda(r,t)
=-\frac12\int\mathcal D t\,H_2\bigl(\psi_\lambda(r,t)\bigr),
\end{aligned}
\label{eq:S-F}
\end{equation}
where $H_2(p)=-p\ln p-(1-p)\ln(1-p)$, and the last equality again
uses the symmetry under $t\mapsto-t$. This is Eq.~\eqref{eq:F}.
Within the RS variational prescription, equilibrium is selected by
the global maximum,
\begin{equation}
\phi_{\rm RS}(\alpha,\lambda)
=\max_{0\leq r<1}\mathcal F_\lambda(r;\alpha),
\qquad
r_\star\in\operatorname*{arg\,max}_{0\leq r<1}
\mathcal F_\lambda(r;\alpha),
\label{eq:S-bayes-RS-selection}
\end{equation}
not by continuation of a stationary root. This prescription and the
replica calculation remain distinct from the exact finite-$N$
Nishimori identity. Since $\mathcal F_\lambda(0;\alpha)=-\alpha\ln2$,
the geometric term is nonpositive and $H_2\geq0$, the RS result
satisfies $-\alpha\ln2\leq\phi_{\rm RS}\leq0$.

\paragraph{Information carried by the training labels.}
For discrete labels, the conditional Shannon entropy is the average
uncertainty about the label sequence when the inputs are known.
Using natural logarithms, it is measured in nats. By
Eq.~\eqref{eq:S-bayes-evidence},
\begin{equation}
\begin{aligned}
H(\bm\xi\mid X)
&=-\E_X\sum_{\bm\eta\in\{\pm1\}^P}
V_N\bigl(\mathcal T_P(X,\bm\eta)\bigr)
\ln V_N\bigl(\mathcal T_P(X,\bm\eta)\bigr)\\
&=-\E\ln V_N(\mathcal T_P).
\end{aligned}
\label{eq:S-evidence-entropy}
\end{equation} The second equality averages the log-volume over
the labels actually generated by the teacher, with their conditional
probabilities from Eq.~\eqref{eq:S-bayes-evidence}.

The conditional mutual information measures how much observing the
labels reveals about the teacher when $X$ is known. Its entropy
identity is
\begin{equation}
I(\bw^*;\bm\xi\mid X)
=H(\bm\xi\mid X)-H(\bm\xi\mid\bw^*,X)
=H(\bm\xi\mid X),
\label{eq:S-information-identity}
\end{equation}
because the labels are determined by $(\bw^*,X)$. Introducing the
finite-size counterpart of Eq.~\eqref{eq:S-phi},
\begin{equation}
\phi_{N,P}(\lambda)
\coloneqq\frac1N\E\ln V_N(\mathcal T_P)
=-\frac1N H(\bm\xi\mid X)
=-\frac1N I(\bw^*;\bm\xi\mid X),
\label{eq:S-finite-information}
\end{equation}
we obtain, in the proportional limit $P/N\to\alpha$,
\begin{equation}
\phi(\alpha,\lambda)
=-\lim_{\substack{N\to\infty\\P/N\to\alpha}}
\frac1N I(\bw^*;\bm\xi\mid X),
\qquad
-\alpha\ln2\leq\phi(\alpha,\lambda)\leq0,
\qquad \phi(0,\lambda)=0,
\label{eq:S-information-limit}
\end{equation}
assuming the limit exists. The bounds follow from
$0\leq I(\bw^*;\bm\xi\mid X)\leq P\ln2$, since each binary label
carries at most $\ln2$ nats. Thus $-\phi$ measures the information
about the teacher supplied by the labels per weight. The
nonpositivity of $\phi$ is consistent with its definition as the
logarithm of a normalized volume; the associated Shannon entropy
and mutual information are nonnegative.

\paragraph{Prediction of a fresh label and Gibbs generalization.}
Let $\bx^{P+1}$ be a fresh input, independent of the teacher and of
the training inputs, and let
$\xi^{P+1}=g_{\bw^*}(\bx^{P+1})$ be its teacher-generated label.
The posterior predictive probability of a positive label is
\begin{equation}
\pi_{\mathcal T_P}(\bx)
\coloneqq\int\de\mu_{\mathcal T_P}(\bw)\,
\one\{g_\bw(\bx)=+1\},
\qquad
\PP(\xi^{P+1}=+1\mid\mathcal T_P,\bx^{P+1})
=\pi_{\mathcal T_P}(\bx^{P+1}).
\label{eq:S-posterior-predictive-probability}
\end{equation}
Here $\pi_{\mathcal T_P}$ depends on the actual training set; it is
distinct from the scalar Gaussian probability $\psi_\lambda$ used
in the RS calculation. Its average binary entropy gives the
uncertainty about the fresh label:
\begin{equation}
H(\xi^{P+1}\mid\mathcal T_P,\bx^{P+1})
=\E_{\mathcal T_P,\bx^{P+1}}
H_2\bigl(\pi_{\mathcal T_P}(\bx^{P+1})\bigr).
\label{eq:S-finite-predictive-entropy}
\end{equation}
This conditional entropy already includes the average over the
variables after the conditioning bar; it is a number, rather than
an entropy for one fixed training set.

The chain rule makes its relation to the information gained from
one more example explicit:
\begin{equation}
I\bigl(\bw^*;(\bm\xi,\xi^{P+1})\mid X,\bx^{P+1}\bigr)
-I(\bw^*;\bm\xi\mid X)=I(\bw^*;\xi^{P+1}\mid\mathcal T_P,\bx^{P+1})=H(\xi^{P+1}\mid\mathcal T_P,\bx^{P+1}).
\label{eq:S-information-chain-rule}
\end{equation}
In the first equality, independence of $\bx^{P+1}$ means that
conditioning additionally on this input does not change the
information in the first $P$ labels. In the second, the fresh label
is determined once the teacher and its input are known. Combining
Eq.~\eqref{eq:S-information-chain-rule} with
Eq.~\eqref{eq:S-finite-information} gives the exact finite-size
increment
\begin{equation}
-N\bigl[\phi_{N,P+1}(\lambda)-\phi_{N,P}(\lambda)\bigr]
=H(\xi^{P+1}\mid\mathcal T_P,\bx^{P+1}).
\label{eq:S-predictive-increment}
\end{equation}

To pass from this increment to a derivative, note that conditioning
on another training example can only reduce the average predictive
entropy. Exchangeability of fresh examples therefore gives
\begin{equation}
\begin{aligned}
H(\xi^{P+2}\mid\mathcal T_{P+1},\bx^{P+2})
&\leq H(\xi^{P+2}\mid\mathcal T_P,\bx^{P+2})\\
&=H(\xi^{P+1}\mid\mathcal T_P,\bx^{P+1}).
\end{aligned}
\end{equation}
Consequently, the piecewise-linear interpolation of
$\phi_{N,P}(\lambda)$ in the load $P/N$ is convex. At every
differentiability point $\alpha>0$ of its limiting function, the
slopes converge to the limiting derivative: each slope is bounded
between the neighboring secant slopes, whose limits approach that
derivative. Thus
\begin{equation}
-\partial_\alpha\phi(\alpha,\lambda)
=\lim_{\substack{N\to\infty\\P/N\to\alpha}}
H(\xi^{P+1}\mid\mathcal T_P,\bx^{P+1}).
\label{eq:S-predictive}
\end{equation}
This is the predictive-entropy interpretation used in the main text.

For the RS prediction, differentiating the maximizing value in Eq.~\eqref{eq:S-bayes-RS-selection} along a smooth equilibrium
branch gives
\begin{equation}
-\partial_\alpha\phi_{\rm RS}(\alpha,\lambda)
=\int\mathcal D t\,
H_2\bigl(\psi_\lambda(r_\star(\alpha,\lambda),t)\bigr).
\label{eq:S-predictive-RS}
\end{equation}
The derivative through $r_\star$ vanishes by stationarity. At
coexistence, the two one-sided derivatives are evaluated at the
corresponding maximizing overlaps. Their difference gives the RS
prediction for the drop in predictive uncertainty discussed in
Appendix~\ref{sec:structure}. Equations~\eqref{eq:S-finite-information}
and~\eqref{eq:S-predictive} concern the exact model, whereas
Eq.~\eqref{eq:S-predictive-RS} is its replica-symmetric prediction.

Finally, conditioned on $\mathcal T_P$ and a fresh input, the
teacher and a Gibbs student give independent labels, each positive
with probability $\pi_{\mathcal T_P}(\bx)$. Their disagreement
probability is therefore $2\pi_{\mathcal T_P}(\bx)
[1-\pi_{\mathcal T_P}(\bx)]$, so
\begin{equation}
\E\,\varepsilon_{g,N}
=2\E_{\mathcal T_P,\bx^{P+1}}
\left[\pi_{\mathcal T_P}(\bx^{P+1})
\bigl(1-\pi_{\mathcal T_P}(\bx^{P+1})\bigr)\right].
\label{eq:S-Bayesian-Gibbs-risk}
\end{equation}
This is the error of a single posterior sample, averaged over
the planted data. The Bayes-optimal classifier for binary
classification instead chooses the more probable label and has
conditional error $\min\{\pi_{\mathcal T_P}(\bx),
1-\pi_{\mathcal T_P}(\bx)\}$. Thus the matched Bayesian setting does
not identify the Gibbs error with the error of the posterior
majority vote; the learning curves in this paper concern the
former.

\begin{remark}[Relation to generalized linear models]\label{rem.dist_lim}
{
For a Gaussian prior $\mathcal{N}(0,1)$ on the weights, the Bayes-optimal replica formula for generalized linear models~\cite{Barbier2019} contains the prior contribution $\inf_{\hat{r}\geq0}\left\{\frac{1}{2}\big[\hat{r}-\ln(1+\hat{r})\big]-\frac{1}{2}\hat{r}r\right\}$. For $0\leq r<1$, the minimum is attained at $\hat{r}=r/(1-r)$ and equals $\frac{1}{2}[\ln(1-r)+r]$.  The channel term for $y=s_\lambda(z)$ equals $2\mathcal{I}_\lambda(r)$.
} Hence Eq.~\eqref{eq:S-F} coincides with that formula, which is a theorem for Gaussian inputs under regularity assumptions on the channel~\cite{Barbier2019}. The deterministic, discontinuous channel $s_\lambda$ and the Rademacher inputs used here are not directly covered.
\end{remark}

\section{Joint distribution of the local fields}
\label{sec:ConvergenceLocalFields}

\subsection{Gaussian limit of the local fields}
\label{sec:GaussianFieldLimit}

The convergence of $\rho_N(\bm{h}|\bm{W})$ to the Gaussian measure~\eqref{eq:GaussianMeasureCorrelated} is pretty standard in the literature~\cite{watkin1993,engel2001}. We present here for completeness a self-sufficient re-derivation.
Fix the number $n$ of replicas and a sequence of weight vectors whose
overlap matrices converge to $Q$. Write
$H_a(\bx\mid\bW)=\bW^a\cdot\bx/\sqrt N$ for the random local
fields. Let us consider the characteristic function
\begin{equation}
    \varphi_N(\bm{k}|\bm{W})\coloneqq \int\dd\bm{h}\, {\rm e}^{\ii \bm{k}\cdot\bm{h}}\rho_N(\bm{h}|\bm{W}),
\end{equation}
which can be computed explicitly using \eqref{eq:induced-field-distribution}:
\begin{align}
\varphi_N(\bm{k}|\bm{W})&= \E_{\bm x}
\int_{\mathbbm{R}^{n+1}}\dd \bm{h}\,
{\rm e}^{\ii\bm k\cdot\bm h}
\prod_{a=0}^{n}\delta\big(h_a-H_a(\bm x|\bm{W})\big) \notag\\
&= \E_{\bm x}{\rm e}^{\ii\bm k\cdot\bm H(\bm x|\bm{W})}=
\E_{\bm x}\exp\Big(\frac{\ii}{\sqrt N}\sum_{j=1}^{N}x_j\sum_{a=0}^{n}k_aW_j^a\Big) \notag\\
&= \prod_{j=1}^{N}\cos\Big(\frac{1}{\sqrt N}\sum_{a=0}^{n}k_a W_j^a\Big).
\end{align}
Then the induced distribution of the local fields can be obtained with a Fourier inversion:
\begin{equation}\label{eq:rho-Fourier-inversion}
\rho_N(\bm h\mid \bm{W})=\int_{\mathbbm{R}^{n+1}}
\frac{\dd \bm{k}}{(2\pi)^{n+1}}\,{\rm e}^{-\ii\bm k\cdot\bm h}\,\varphi_N(\bm k|\bm{W}).
\end{equation}
The high-dimensional limit $N\rightarrow\infty$ is easily performed on the level of the characteristic function. In fact, using 
\(\log\cos y=-y^2/2+O(y^4)\) yields
\begin{align}\label{eq:field-characteristic-limit}
\log\varphi_N(\bm k|\bm{W}) &= -\frac{1}{2N}
\sum_{j=1}^{N}\Big(\sum_{a=0}^{n}k_aW_j^a\Big)^2 + o(1) \notag\\
&= -\frac12\sum_{a,b=0}^{n}k_aQ_{ab}k_b+o(1).
\end{align}
The remainder vanishes under the condition that no individual coordinate carries a macroscopic fraction of the norm, i.e.
\begin{equation}
\max_{1\leq j\leq N}\frac{|W_j^a|}{\sqrt N}\longrightarrow0 \qquad \text{as }N\rightarrow\infty, \text{ for every}
\quad a=0,\ldots,n.
\label{eq:replicated-delocalization}
\end{equation} 
This holds with overwhelming probability for a fixed number of vectors drawn from the spherical measure.  We therefore obtain
\begin{equation}\label{eq:Gaussian-characteristic-function}
\varphi_N(\bm k) \longrightarrow \exp\!\left(-\frac12\bm k^{\mathsf T}Q\bm k\right).
\end{equation}
L\'evy's continuity theorem therefore gives weak convergence of the
field distributions:
\begin{equation}
\rho_N(\bm h\mid\bW)\,\de\bm h
\;\longrightarrow\;\de\gamma_Q(\bm h),
\qquad \text{as }N\to\infty.
\label{eq:S-field-weak-limit}
\end{equation}
For positive-definite $Q$, Fourier inversion of the limiting
characteristic function gives the density in
Eq.~\eqref{eq:GaussianMeasureCorrelated}. For singular $Q$, the
limiting Gaussian law is still defined by its characteristic function.
The convergence is understood in the sense of measures, rather than
pointwise convergence of densities: at finite $N$, the Rademacher
patterns produce a discrete field distribution. The discontinuities
of the one-pattern integrand lie on the hyperplanes
$h_a=k\pi/\lambda$. Each has zero probability under the limiting
Gaussian law because every field has unit variance. Thus weak
convergence also implies the convergence of the pattern average
in Eq.~\eqref{eq:S-Zn}.
\subsection{Gaussian decomposition under replica symmetry}
\label{sec:GaussianDecomposition}

We now derive explicitly the representation~\eqref{eq:S-rep} used to
evaluate the one-pattern contribution in Appendix~\ref{sec:RS-ansatz}. We work with the
limiting Gaussian fields and assume $0\leq r^2\leq q<1$, the domain in
which the decomposition below uses real Gaussian variables. The
replica-symmetric covariance matrix is
\begin{equation}
Q_{\mathrm{RS}}=
\begin{pmatrix}
1 & r\one_n^{\mathsf T}\\
r\one_n & (1-q)\mathbb I_n+q\one_n\one_n^{\mathsf T}
\end{pmatrix},
\label{eq:S-QRS-block}
\end{equation}
where $\one_n=(1,\ldots,1)^{\mathsf T}$ and $\mathbb I_n$ is the
$n\times n$ identity matrix. To tackle the correlations between the
student replicas and the teacher, note that this matrix can be viewed
as the sum of three contributions:
\begin{equation}
Q_{\mathrm{RS}}=\bm v_0\bm v_0^{\mathsf T}
+\bm v_{\mathrm c}\bm v_{\mathrm c}^{\mathsf T}
+(1-q)\Pi_{\mathrm s},
\label{eq:S-QRS-rank-decomposition}
\end{equation}
with
\begin{equation}
\bm v_0=
\begin{pmatrix}1\\r\one_n\end{pmatrix},
\qquad
\bm v_{\mathrm c}=
\begin{pmatrix}0\\\sqrt{q-r^2}\,\one_n\end{pmatrix},
\qquad
\Pi_{\mathrm s}=
\begin{pmatrix}0&0\\0&\mathbb I_n\end{pmatrix}.
\end{equation}
The three terms have the following interpretation: the first produces
the teacher field and the part of every student field correlated with
it; the second is an additional fluctuation \emph{shared} by all
student replicas which is not accounted for by their common
correlation with the teacher; the last contains the independent
fluctuations specific to each replica. In particular, when $q>r^2$,
the rank-one contribution $\bm v_{\mathrm c}\bm v_{\mathrm c}^{\mathsf T}$
couples the student fields even after conditioning on the teacher
field $h_0$.

Indeed, the conditional mean and covariance of the student fields are
\begin{equation}
\E[h_a\mid h_0]=rh_0,
\qquad
\operatorname{Cov}(h_a,h_b\mid h_0)
=(1-q)\delta_{ab}+q-r^2,
\qquad a,b=1,\ldots,n.
\end{equation}
Thus their conditional characteristic function, with
$\bm k=(k_1,\ldots,k_n)$, is
\begin{equation}
\widehat\rho_{\mathrm{RS}}(\bm k\mid h_0)
\coloneqq
\E\!\left[\exp\!\left(\ii\sum_{a=1}^n k_a h_a\right)\middle|h_0\right]
=\exp\!\left(
\ii rh_0\sum_{a=1}^n k_a
-\frac{1-q}{2}\sum_{a=1}^n k_a^2
-\frac{q-r^2}{2}\Bigl(\sum_{a=1}^n k_a\Bigr)^2
\right).
\label{eq:S-conditional-characteristic}
\end{equation}
The residual coupling occurs through the single collective coordinate
$\sum_{a=1}^n k_a$. It can be removed at the cost of introducing an
auxiliary normal random variable via the Hubbard--Stratonovich identity
\begin{equation}
{\rm e}^{-cX^2/2}
=\int\mathcal D\tau\,{\rm e}^{\ii\sqrt c\,\tau X},
\qquad c\geq0,
\qquad \mathcal D\tau=\frac{{\rm e}^{-\tau^2/2}}{\sqrt{2\pi}}\,\de\tau.
\label{eq:S-HS-Fourier}
\end{equation}
Applying this identity with $c=q-r^2$ yields
\begin{equation}
\widehat\rho_{\mathrm{RS}}(\bm k\mid h_0)
=\int\mathcal D\tau\prod_{a=1}^{n}
\exp\Bigl([\ii k_a\bigl(rh_0+\sqrt{q-r^2}\,\tau\bigr)
-\frac{1-q}{2}k_a^2\Bigr).
\label{eq:S-conditional-characteristic-factorized}
\end{equation}
Each factor is the characteristic function of a Gaussian variable
with mean $rh_0+\sqrt{q-r^2}\,\tau$ and variance $1-q$. Fourier
inversion therefore gives the conditional density
\begin{equation}
\rho_{\mathrm{RS}}(h_1,\ldots,h_n\mid h_0)
=\int\mathcal D\tau\prod_{a=1}^{n}
\frac{1}{\sqrt{2\pi(1-q)}}
\exp\Bigl(-\frac{\bigl(h_a-rh_0-\sqrt{q-r^2}\,\tau\bigr)^2}{2(1-q)}\Bigr).
\label{eq:S-conditional-density-fields}
\end{equation}
This shows that the student fields are independent conditionally on
both $h_0$ and the auxiliary variable $\tau$. Equivalently, we can
introduce independent standard Gaussians $h_0,\tau,z_1,\ldots,z_n$,
each with law $\mathcal N(0,1)$, and represent the student fields as
\begin{equation}
h_a=rh_0+\sqrt{q-r^2}\,\tau+\sqrt{1-q}\,z_a,
\qquad a=1,\ldots,n.
\label{eq:S-fields-from-HS}
\end{equation}

To use this decomposition in the pattern contribution, recall that
all student labels must agree with the teacher label. Separating the
two possible labels gives, away from the zero-probability decision
boundaries,
\begin{equation}
\prod_{a=1}^n\Theta\bigl(\sin(\lambda h_0)\sin(\lambda h_a)\bigr)=\Theta\bigl(\sin(\lambda h_0)\bigr)
\prod_{a=1}^n\Theta\bigl(\sin(\lambda h_a)\bigr)+\Theta\bigl(-\sin(\lambda h_0)\bigr)
\prod_{a=1}^n\Theta\bigl(-\sin(\lambda h_a)\bigr).
\end{equation}
The centered Gaussian measure is invariant under simultaneous sign
reversal $(h_0,\ldots,h_n)\mapsto(-h_0,\ldots,-h_n)$, so the two
terms have equal integrals. Consequently,
\begin{equation}
\mathcal Z_{n,\lambda}(q,r)
=2\int\de\gamma_{Q_{\mathrm{RS}}}(\bh)\,
\Theta\bigl(\sin(\lambda h_0)\bigr)
\prod_{a=1}^n\Theta\bigl(\sin(\lambda h_a)\bigr).
\label{eq:S-Zn-positive-label}
\end{equation}
Using Eq.~\eqref{eq:S-fields-from-HS}, this becomes
\begin{equation}
\mathcal Z_{n,\lambda}(q,r)=2\int\mathcal D h_0\,\mathcal D\tau\,\Theta\bigl(\sin(\lambda h_0)\bigr)\left[\int\mathcal D z\, \Theta\Bigl(\sin\bigl(\lambda(rh_0+\sqrt{q-r^2}\,\tau+\sqrt{1-q}\,z)\bigr)\Bigr)\right]^{n}.
\label{eq:S-Zn-h0-tau}
\end{equation}
The power $n$ appears because the replicas become independent once the
two common variables $h_0$ and $\tau$ are fixed.

The common part of every student field is
$rh_0+\sqrt{q-r^2}\,\tau$, whose variance is
\begin{equation}
\E\bigl[(rh_0+\sqrt{q-r^2}\,\tau)^2\bigr]
=r^2+(q-r^2)=q.
\end{equation}
For $q>0$, normalize this common mode by defining
$\sqrt q\,t=rh_0+\sqrt{q-r^2}\,\tau$. To complete the change of
variables, introduce the orthogonal Gaussian mode $y$ through
\begin{equation}
\begin{pmatrix}t\\y\end{pmatrix}
=\begin{pmatrix}
r/\sqrt q & \sqrt{q-r^2}/\sqrt q\\[1mm]
\sqrt{q-r^2}/\sqrt q & -r/\sqrt q
\end{pmatrix}
\begin{pmatrix}h_0\\\tau\end{pmatrix}.
\label{eq:S-orthogonal-rotation}
\end{equation}
The matrix is orthogonal, so $t$ and $y$ are again independent
$\mathcal N(0,1)$ variables, and
$\mathcal D h_0\,\mathcal D\tau=\mathcal D t\,\mathcal D y$.
Inverting the transformation gives
\begin{equation}
h_0=\frac{rt}{\sqrt q}+\sqrt{1-\frac{r^2}{q}}\,y,
\qquad
h_a=\sqrt q\,t+\sqrt{1-q}\,z_a.
\label{eq:S-rotated-fields}
\end{equation}
This is Eq.~\eqref{eq:S-rep}. It is especially convenient because,
conditionally on the single common variable $t$, the teacher field
and all the student fields are mutually independent. In these
variables Eq.~\eqref{eq:S-Zn-h0-tau} reads
\begin{equation}
\mathcal Z_{n,\lambda}(q,r)=2\int\mathcal D t\,
\int\mathcal D y\,
\Theta\biggl(\sin\biggl(\lambda\Bigl(
\frac{rt}{\sqrt q}+\sqrt{1-\frac{r^2}{q}}\,y\Bigr)\biggr)\biggr) \left[\int\mathcal D z\,\Theta\Bigl(\sin\bigl(\lambda(
\sqrt q\,t+\sqrt{1-q}\,z)\bigr)\Bigr)\right]^{n}.
\label{eq:S-Zn-t-y}
\end{equation}
The integrals over $y$ and $z$ are precisely the Gaussian convolutions
defined in Eq.~\eqref{eq:S-p}:
\begin{equation}
\chi_\lambda(q,r,t)
=p_\lambda\!\left(\frac{rt}{\sqrt q},1-\frac{r^2}{q}\right),
\qquad
\psi_\lambda(q,t)=p_\lambda(\sqrt q\,t,1-q).
\end{equation}
Thus $\chi_\lambda$ is the probability of a positive teacher label
conditioned on $t$, and $\psi_\lambda$ is the corresponding probability
for each student replica. We recover Eq.~\eqref{eq:S-ZRS}:
\begin{equation}
\mathcal Z_{n,\lambda}(q,r)
=2\int\mathcal D t\,\chi_\lambda(q,r,t)\psi_\lambda(q,t)^n.
\end{equation}
When $q=r^2>0$, the teacher has no residual fluctuation at fixed $t$;
the expression for $\chi_\lambda$ is then understood using
$p_\lambda(\mu,0)=\Theta(\sin\lambda\mu)$. At $q=0$ the condition
$r^2\leq q$ forces $r=0$. All original fields are then independent,
and the pattern factor is directly $\mathcal Z_{n,\lambda}(0,0)=2^{-n}$,
without introducing the normalized common mode $t$.

%=====================================================================
\section{Generalization error and asymptotic analysis}
\label{sec:generr}
\label{sec:asymptotics}
%=====================================================================

We first discuss the small- and large-frequency limits of the reduced
RS potential, then derive the generalization error and its asymptotic
bounds. Finally, we analyze the response at small load and the
stationary branches at large frequency or large load, using the
stationarity and Fisher-information representations derived in
Appendix~\ref{sec:structure}.

\subsection{Classical limit \texorpdfstring{$\lambda\to0^+$}{lambda to zero}}

For $z\ne0$ and $0<\lambda<\pi/|z|$,
$\Theta(\sin\lambda z)=\Theta(z)$. Dominated convergence therefore
gives, for $v>0$,
\begin{equation}
\lim_{\lambda\to0^+}p_\lambda(\mu,v)
=\E_{Z\sim\mathcal N(\mu,v)}[\Theta(Z)]
=\Phi\Bigl(\frac{\mu}{\sqrt v}\Bigr),
\label{eq:S-small-lambda-probability}
\end{equation}
where, as in Appendix~\ref{sec:circuit},
\begin{equation}
\Phi(x)=\int_{-\infty}^{x}\mathcal D z
=\frac12\biggl(1+\operatorname{erf}\Bigl(\frac{x}{\sqrt2}\Bigr)\biggr)
\end{equation}
is the standard normal distribution cumulative function. For $v=0$ and
$\mu\ne0$, the limit is $\Theta(\mu)$. For fixed $0\leq r<1$,
Eq.~\eqref{eq:S-small-lambda-probability} gives
\begin{equation}
\lim_{\lambda\to0^+}\psi_\lambda(r,t)
=\Phi\!\left(\sqrt{\frac{r}{1-r}}\,t\right).
\label{eq:S-classical-psi}
\end{equation}
Consequently, the integrand in the replica-symmetric potential
\eqref{eq:S-F} converges pointwise to the one corresponding to the
classical spherical teacher--student perceptron. Since the binary
entropy satisfies $0\leq H_2(p)\leq\ln2$, dominated convergence also
applies to the $t$-integral. Thus, for fixed $r<1$ and finite $\alpha$,
\begin{equation}
\lim_{\lambda\to0^+}\mathcal F_\lambda(r;\alpha)
=\frac12\ln(1-r)+\frac r2
-\alpha\int\mathcal D t\,
H_2\!\left[\Phi\!\left(\sqrt{\frac{r}{1-r}}\,t\right)\right],
\label{eq:S-classical-potential}
\end{equation}
which is the standard RS free entropy functional of the spherical
teacher--student perceptron~\cite{Seung1992,engel2001}. In this limit,
the linear-response coefficient defined in Eq.~\eqref{eq:S-kappa}
satisfies $\kappa_\lambda\to(2\pi)^{-1/2}$.

\subsection{Large-frequency limit \texorpdfstring{$\lambda\to\infty$}{lambda to infinity}}

\begin{prop}\label{prop:psi-inf}
For every $0\leq r<1$ and $\lambda>0$,
\begin{equation}
\left|\psi_\lambda(r,t)-\frac12\right|
\leq\frac{1}{2\lambda}\sqrt{\frac{\pi}{2(1-r)}}.
\label{eq:S-psi-inf-bound}
\end{equation}
In particular, for every fixed $r<1$,
\begin{equation}
\lim_{\lambda\to\infty}\psi_\lambda(r,t)=\frac12
\end{equation}
uniformly in $t\in\mathbb R$. The convergence is also uniform in
$(r,t)\in[0,\bar r]\times\mathbb R$ for every $\bar r<1$.
\end{prop}

\begin{proof}
The estimate follows from an integration by parts. From the definition
of $\psi_\lambda$,
\begin{equation}
\begin{aligned}
\psi_\lambda(r,t)-\frac12
&=\frac{1}{\sqrt{2\pi(1-r)}}\int_{\mathbb R}\de x\,
\exp\Bigl(-\frac{(x-\sqrt r\,t)^2}{2(1-r)}\Bigr)
\Theta(\sin\lambda x)-\frac12\\
&=\frac{1}{\sqrt{2\pi(1-r)}}\int_{\mathbb R}\de x\,
\exp\Bigl(-\frac{(x-\sqrt r\,t)^2}{2(1-r)}\Bigr)g(\lambda x),
\end{aligned}
\end{equation}
where we used the Gaussian normalization to bring $1/2$ inside the
integral and introduced the centered square wave
$g(x)=\Theta(\sin x)-1/2$.

Since $g$ has zero average over a period, it has a bounded periodic
primitive. Specifically, we choose
\begin{equation}
\widetilde G(x)=
\begin{cases}
\dfrac{x}{2}-\dfrac{\pi}{4},&0\leq x\leq\pi,\\[2mm]
\dfrac{3\pi}{4}-\dfrac{x}{2},&\pi\leq x\leq2\pi,
\end{cases}
\label{eq:S-periodic-primitive}
\end{equation}
extended periodically outside $[0,2\pi]$. This is the triangular
primitive $\int_0^x g(u)\,\de u$ shifted by $-\pi/4$, so that it has
zero mean and $\|\widetilde G\|_\infty=\pi/4$. Since
\begin{equation}
g(\lambda x)=\frac1\lambda\frac{\de}{\de x}\widetilde G(\lambda x)
\end{equation}
almost everywhere, integration by parts gives
\begin{equation}
\psi_\lambda(r,t)-\frac12
=\frac{1}{\lambda\sqrt{2\pi(1-r)}}
\int_{\mathbb R}\de x\,
\frac{x-\sqrt r\,t}{1-r}
\exp\Bigl(-\frac{(x-\sqrt r\,t)^2}{2(1-r)}\Bigr)
\widetilde G(\lambda x).
\end{equation}
The boundary term vanishes because $\widetilde G$ is bounded and thev Gaussian factor tends to zero. Therefore
\begin{equation}
\begin{aligned}
\left|\psi_\lambda(r,t)-\frac12\right|
&\leq\frac{\|\widetilde G\|_\infty}{\lambda(1-r)}
\frac{1}{\sqrt{2\pi(1-r)}}
\int_{\mathbb R}\de x\,|x-\sqrt r\,t|
\exp\Bigl(-\frac{(x-\sqrt r\,t)^2}{2(1-r)}\Bigr)\\
&=\frac{\pi}{4\lambda(1-r)}\sqrt{\frac{2(1-r)}{\pi}}
=\frac{1}{2\lambda}\sqrt{\frac{\pi}{2(1-r)}}.
\end{aligned}
\end{equation}
This proves Eq.~\eqref{eq:S-psi-inf-bound}. The right-hand side is independent of $t$ and tends to zero uniformly for $0\leq r\leq\bar r<1$.
\end{proof}

\begin{prop}[Large frequency at fixed load]\label{prop:large-lambda}
Let $r_\star(\alpha,\lambda)$ denote any maximizer of
$\mathcal F_\lambda(r;\alpha)$ in Eq.~\eqref{eq:S-F}. Then, for every
fixed finite $\alpha\geq0$,
\begin{equation}
r_\star(\alpha,\lambda)\longrightarrow0,
\qquad
\phi_{\rm RS}(\alpha,\lambda)\longrightarrow-\alpha\ln2,
\qquad
\varepsilon_\lambda\bigl(r_\star(\alpha,\lambda)\bigr)
\longrightarrow\frac12
\quad\text{as }\lambda\to\infty.
\end{equation}
The limiting RS potential has the unique maximizer $r=0$; hence the
limit of $r_\star$ is independent of the choice of maximizer at finite
$\lambda$.
\end{prop}

\begin{proof}
We need uniform convergence on an interval containing all maximizers.
The first step is therefore to prove that the maximizers stay bounded
away from $1$ uniformly in $\lambda$. Recall that the pattern
contribution to $\mathcal F_\lambda$ is nonpositive, since
$H_2\geq0$. Hence, for every $r\in[0,1)$,
\begin{equation}
\mathcal F_\lambda(r;\alpha)
\leq\frac12\bigl[\ln(1-r)+r\bigr].
\end{equation}
If $r_\star$ is a maximizer, then
$\mathcal F_\lambda(r_\star;\alpha)\geq
\mathcal F_\lambda(0;\alpha)=-\alpha\ln2$, and therefore
\begin{equation}
-\alpha\ln2
\leq\frac12\bigl[\ln(1-r_\star)+r_\star\bigr]
\leq\frac12\bigl[1+\ln(1-r_\star)\bigr].
\end{equation}
Thus every maximizer satisfies
\begin{equation}
1-r_\star\geq\delta_\alpha,
\qquad
\delta_\alpha\coloneqq\exp[-1-2\alpha\ln2]
={\rm e}^{-1}2^{-2\alpha}>0.
\label{eq:S-maximizer-compactness}
\end{equation}
For fixed finite $\alpha$, all maximizers are therefore confined to
the compact interval $[0,1-\delta_\alpha]$, independently of $\lambda$.
Continuity of $\mathcal F_\lambda$ on this interval and its divergence
to $-\infty$ as $r\to1$ also ensure that a maximizer exists.

On this interval, Proposition~\ref{prop:psi-inf} implies that
$\psi_\lambda\to1/2$ uniformly in both $r$ and $t$. By continuity of
$H_2$ at $1/2$, and since $\int\mathcal D t=1$, the potential
$\mathcal F_\lambda(r;\alpha)$ converges uniformly to
\begin{equation}
\mathcal F_\infty(r;\alpha)
=\frac12\bigl[\ln(1-r)+r\bigr]-\alpha\ln2.
\label{eq:S-limiting-potential}
\end{equation}
The limiting functional has the unique maximizer $r=0$, since
$\ln(1-r)+r\leq0$, with equality only at $r=0$.

To see explicitly why the maximizers converge, fix any
$0<\eta<1-\delta_\alpha$. For $r\geq\eta$ in the compact interval,
\begin{equation}
\mathcal F_\infty(0;\alpha)-\mathcal F_\infty(r;\alpha)
\geq-\frac12\bigl[\ln(1-\eta)+\eta\bigr]>0.
\end{equation}
Once the uniform difference between $\mathcal F_\lambda$ and
$\mathcal F_\infty$ is smaller than half this positive gap, no such
$r$ can maximize $\mathcal F_\lambda$. Since $\eta$ can be arbitrarily
small, $r_\star(\alpha,\lambda)\to0$. Moreover,
\begin{equation}
\left|\max_{0\leq r\leq1-\delta_\alpha}\mathcal F_\lambda(r;\alpha)
-\max_{0\leq r\leq1-\delta_\alpha}\mathcal F_\infty(r;\alpha)\right|
\leq\sup_{0\leq r\leq1-\delta_\alpha}
\left|\mathcal F_\lambda(r;\alpha)-\mathcal F_\infty(r;\alpha)\right|
\longrightarrow0,
\end{equation}
which proves $\phi_{\rm RS}\to-\alpha\ln2$.

Finally, the error bound~\eqref{eq:S-large-lambda}, proved below,
applies uniformly on the same compact interval. In particular,
\begin{equation}
0\leq\frac12-\varepsilon_\lambda(r_\star)
\leq\frac{4}{\pi^2}
\frac{{\rm e}^{-\lambda^2\delta_\alpha}}
{\bigl(1-{\rm e}^{-4\lambda^2\delta_\alpha}\bigr)^2}
\longrightarrow0.
\end{equation}
This justifies evaluating the large-frequency limit at the
$\lambda$-dependent equilibrium overlap.
\end{proof}

\subsection{Generalization error}

\begin{prop}\label{lem:eps}
For $0\leq r<1$ and $\lambda>0$, the function $\varepsilon_\lambda(r)$
defined by Eq.~\eqref{eq:S-epsdef} has the equivalent representation
\begin{equation}
\varepsilon_\lambda(r)=2\int\mathcal D t\,\psi_\lambda(r,t)\bigl(1-\psi_\lambda(r,t)\bigr),
\label{eq:S-eps1}
\end{equation}
and admits the absolutely convergent series representation
\begin{equation}
\varepsilon_\lambda(r)
=\frac12-\frac{8}{\pi^2}
\sum_{m,n=0}^{\infty}
\frac{\sinh\bigl(\lambda^2r(2m+1)(2n+1)\bigr)}
{(2m+1)(2n+1)}\exp\Bigl(-\frac{\lambda^2}{2}
\bigl((2m+1)^2+(2n+1)^2\bigr)\Bigr).
\label{eq:S-eps-series}
\end{equation}
\end{prop}

\begin{proof}
By Eq.~\eqref{eq:S-epsdef},
\begin{equation}
\varepsilon_\lambda(r)
=\PP\!\left(s_\lambda(U)\neq s_\lambda(V)\right),
\label{eq:S-eps-UV}
\end{equation}
where we may take
\begin{equation}
U=Z,\qquad V=rZ+\sqrt{1-r^2}\,Y,
\label{eq:S-UV-definition}
\end{equation}
with $Y,Z$ independent $\mathcal N(0,1)$ variables. These are centered
and jointly Gaussian fields with
\begin{equation}
\E[U^2]=\E[V^2]=1,\qquad\E[UV]=r.
\end{equation}
For $0\leq r<1$, the same Gaussian pair can be represented more
conveniently as
\begin{equation}
U=\sqrt r\,T+\sqrt{1-r}\,Y,
\qquad
V=\sqrt r\,T+\sqrt{1-r}\,Z,
\label{eq:common-mode}
\end{equation}
where $T,Y,Z$ are independent $\mathcal N(0,1)$ variables. This is the
two-field form of the decomposition in
Appendix~\ref{sec:GaussianDecomposition}, on the line $q=r$.
The advantage of Eq.~\eqref{eq:common-mode} is that, conditioned on
$T=t$, the two fields are independent and identically distributed
according to $\mathcal N(\sqrt r\,t,1-r)$. Hence, by the definition
of $\psi_\lambda$ in Eq.~\eqref{eq:psi},
\begin{equation}
\PP\!\left(s_\lambda(U)=+1\mid T=t\right)
=\PP\!\left(s_\lambda(V)=+1\mid T=t\right)
=\psi_\lambda(r,t).
\end{equation}
The zeros of $\sin(\lambda U)$ have zero Gaussian probability, so the
conditional probability that the two classifiers disagree is
\begin{equation}
\begin{aligned}
\PP\!\left(s_\lambda(U)\neq s_\lambda(V)\mid T=t\right)
&=\psi_\lambda(r,t)\bigl(1-\psi_\lambda(r,t)\bigr)
+\bigl(1-\psi_\lambda(r,t)\bigr)\psi_\lambda(r,t)\\
&=2\psi_\lambda(r,t)\bigl(1-\psi_\lambda(r,t)\bigr).
\end{aligned}
\end{equation}
Averaging over $t$ proves Eq.~\eqref{eq:S-eps1}.

We next derive the series representation. Since $s_\lambda(U)$ and
$s_\lambda(V)$ take values in $\{-1,+1\}$ almost surely,
\begin{equation}
\varepsilon_\lambda(r)=\frac12\bigl(1-C_\lambda(r)\bigr),
\qquad
C_\lambda(r)\coloneqq\E[s_\lambda(U)s_\lambda(V)].
\label{eq:S-label-correlation}
\end{equation}
Conditional independence in Eq.~\eqref{eq:common-mode} gives
\begin{equation}
C_\lambda(r)
=\int\mathcal D t\,\bigl(2\psi_\lambda(r,t)-1\bigr)^2,
\qquad \varepsilon_\lambda(r)=\frac12-2\int\mathcal D t\,\Bigl(\psi_\lambda(r,t)-\frac12\Bigr)^2.
\label{eq:S-eps-square}
\end{equation}
We use the Fourier series after Gaussian smoothing, so that the
interchange of the series and the integral is justified explicitly.
Equation~\eqref{eq:S-p} yields
\begin{equation}
\psi_\lambda(r,t)-\frac12
=\frac2\pi\sum_{m=0}^{\infty}
\frac{\sin\bigl((2m+1)\lambda\sqrt r\,t\bigr)}{2m+1}
\,\exp\Bigl(-\frac{1}{2}\lambda^2(2m+1)^2(1-r)\Bigr).
\label{eq:S-psi-Fourier}
\end{equation}
For $r<1$, the absolute value of each term is bounded independently
of $t$ by a summable Gaussian sequence. The series therefore
converges absolutely and uniformly in $t$, and its square is an
absolutely and uniformly convergent double series. Since
$\int\mathcal D t=1$, this double series can be integrated term by
term in Eq.~\eqref{eq:S-eps-square}.

For real $A,B$, the Gaussian characteristic function
$\int\mathcal D t\,{\rm e}^{\ii At}={\rm e}^{-A^2/2}$ and trigonometric identities give
\begin{equation}
\begin{aligned}
\int\mathcal D t\,\sin(At)\sin(Bt)
&=\frac12\int\mathcal D t\,
\bigl(\cos((A-B)t)-\cos((A+B)t)\bigr)\\
&=\frac12\Bigl({\rm e}^{-(A-B)^2/2}-{\rm e}^{-(A+B)^2/2}\Bigr)
={\rm e}^{-(A^2+B^2)/2}\sinh(AB).
\end{aligned}
\end{equation}
Taking $A=(2m+1)\lambda\sqrt r$ and $B=(2n+1)\lambda\sqrt r$
and combining this Gaussian factor with the smoothing factors in
Eq.~\eqref{eq:S-psi-Fourier} gives Eq.~\eqref{eq:S-eps-series}.

To verify absolute convergence of the resulting expression
explicitly, use $\sinh x\leq\tfrac12{\rm e}^x$ for $x\geq0$ and
$2(2m+1)(2n+1)\leq(2m+1)^2+(2n+1)^2$. With
$c=\lambda^2(1-r)/2>0$, these imply
\begin{equation}
\begin{aligned}
0\leq{}&\sinh\bigl(\lambda^2r(2m+1)(2n+1)\bigr)
\exp\Bigl(-\frac{\lambda^2}{2}
\bigl((2m+1)^2+(2n+1)^2\bigr)\Bigr)\\
\leq{}&\frac12\,\exp\Bigl(-c\bigl((2m+1)^2+(2n+1)^2\bigr)\Bigr).
\end{aligned}
\end{equation}
Consequently, the absolute values of the summands are bounded by a
summable double sequence:
\begin{equation}
\sum_{m,n=0}^{\infty}
\frac{{\rm e}^{-c((2m+1)^2+(2n+1)^2)}}
{2(2m+1)(2n+1)}
=\frac12\biggl(\sum_{m=0}^{\infty}
\frac{{\rm e}^{-c(2m+1)^2}}{2m+1}\biggr)^2<\infty.
\end{equation}
\end{proof}

\begin{prop}\label{prop:eps-limits}
Let $\varepsilon_\lambda(r)$ be the generalization error defined in
Eq.~\eqref{eq:S-epsdef}, with $0\leq r\leq1$.
\begin{enumerate}
\item In the small-frequency limit,
\begin{equation}
\lim_{\lambda\to0^+}\varepsilon_\lambda(r)
=\frac1\pi\arccos r.
\end{equation}
More precisely, for every $\lambda>0$,
\begin{equation}
\Bigl|\varepsilon_\lambda(r)-\frac1\pi\arccos r\Bigr|
\leq4\Bigl(1-\Phi\bigl(\frac\pi\lambda\bigr)\Bigr)
\leq\frac{4\lambda}{\pi\sqrt{2\pi}}
\exp\Bigl(-\frac{\pi^2}{2\lambda^2}\Bigr),
\label{eq:S-small-lambda}
\end{equation}
uniformly for $r\in[0,1]$.
\item For every $0\leq r<1$ and $\lambda>0$,
\begin{equation}
\frac{4}{\pi^2}{\rm e}^{-\lambda^2(1-r)}
\bigl(1-{\rm e}^{-2\lambda^2r}\bigr)
\leq\frac12-\varepsilon_\lambda(r)
\leq\frac{4}{\pi^2}
\frac{{\rm e}^{-\lambda^2(1-r)}}
{\bigl(1-{\rm e}^{-4\lambda^2(1-r)}\bigr)^2}.
\label{eq:S-large-lambda}
\end{equation}
Therefore, for each fixed $0\leq r<1$,
\begin{equation}
\lim_{\lambda\to\infty}\varepsilon_\lambda(r)=\frac12,
\label{eq:S-large-lambda-limit}
\end{equation}
and the leading correction for fixed $0<r<1$ is
\begin{equation}
\frac12-\varepsilon_\lambda(r)
\sim\frac{4}{\pi^2}{\rm e}^{-\lambda^2(1-r)},
\qquad\lambda\to\infty.
\label{eq:S-large-lambda-sharp}
\end{equation}
At $r=0$ one has $\varepsilon_\lambda(0)=1/2$ for every $\lambda>0$,
whereas at $r=1$ one has $\varepsilon_\lambda(1)=0$. In particular,
the large-$\lambda$ limit is not uniform as $r\to1$.
\end{enumerate}
\end{prop}

\begin{proof}
We begin with the small-$\lambda$ regime. Let $(U,V)$ be the Gaussian
pair in Eq.~\eqref{eq:S-UV-definition} and set
\begin{equation}
a_\lambda=\frac\pi\lambda.
\end{equation}
For $|x|<a_\lambda$, one has
$s_\lambda(x)=\sgn[\sin(\lambda x)]=\sgn x$. Consequently, on the
event $\{|U|<a_\lambda,\ |V|<a_\lambda\}$, the oscillatory
classifier has exactly the same disagreement indicator as the
classical perceptron:
\begin{equation}
\one\{s_\lambda(U)\neq s_\lambda(V)\}
=\one\{UV<0\}
\quad\text{if }|U|<a_\lambda\text{ and }|V|<a_\lambda,
\end{equation}
apart from the zero-probability event $UV=0$. The classical
disagreement probability is
\begin{equation}
\PP(UV<0)=\frac1\pi\arccos r.
\end{equation}
Indeed, writing $r=\cos\vartheta$ in
Eq.~\eqref{eq:S-UV-definition}, the two signs are determined by
projections of the isotropic Gaussian vector $(Z,Y)$ onto directions
separated by the angle $\vartheta$. They disagree in two angular
sectors of total opening $2\vartheta$, giving probability
$\vartheta/\pi$.

The two disagreement events can therefore differ only if at least
one field leaves $(-a_\lambda,a_\lambda)$. Taking expectations of
the indicators and using the union bound gives
\begin{align}\nonumber
\Bigl|\varepsilon_\lambda(r)-\frac1\pi\arccos r\Bigr|
&\leq\PP\bigl(|U|\geq a_\lambda\text{ or }|V|\geq a_\lambda\bigr)\\ \nonumber
&\leq\PP(|U|\geq a_\lambda)+\PP(|V|\geq a_\lambda)\\
&=4\bigl(1-\Phi(a_\lambda)\bigr).
\label{eq:S-small-union-bound}
\end{align}
Here each field has the standard normal marginal, so
\begin{equation}
\PP(|U|\geq a_\lambda)
=\PP(U\geq a_\lambda)+\PP(U\leq-a_\lambda)
=2[1-\Phi(a_\lambda)],
\end{equation}
and similarly for $V$, independently of their correlation. Finally,
the Gaussian tail bound
\begin{equation}
1-\Phi(x)\leq\frac{1}{x\sqrt{2\pi}}{\rm e}^{-x^2/2},
\qquad x>0,
\end{equation}
gives the second inequality in Eq.~\eqref{eq:S-small-lambda} and
proves uniform convergence on $r\in[0,1]$.

We now turn to the large-$\lambda$ regime. By
Proposition~\ref{lem:eps},
\begin{equation}
\frac12-\varepsilon_\lambda(r)
=\frac{8}{\pi^2}\sum_{m,n=0}^{\infty}
\frac{\sinh\bigl(\lambda^2r(2m+1)(2n+1)\bigr)}
{(2m+1)(2n+1)}
\exp\Bigl(-\frac{\lambda^2}{2}
\bigl((2m+1)^2+(2n+1)^2\bigr)\Bigr).
\end{equation}
For $r\geq0$, every term is nonnegative. Retaining only the term
$m=n=0$ gives
\begin{equation}
\frac12-\varepsilon_\lambda(r)
\geq\frac{8}{\pi^2}{\rm e}^{-\lambda^2}\sinh(\lambda^2r)
=\frac{4}{\pi^2}{\rm e}^{-\lambda^2(1-r)}
\bigl(1-{\rm e}^{-2\lambda^2r}\bigr).
\label{eq:S-large-lower-bound}
\end{equation}
For the upper bound, using $\sinh x\leq\tfrac12{\rm e}^x$ for
$x\geq0$ yields
\begin{equation}
\frac12-\varepsilon_\lambda(r)
\leq\frac{4}{\pi^2}\sum_{m,n=0}^{\infty}
\frac{\exp\Bigl(-\frac{\lambda^2}{2}
\bigl((2m+1)^2+(2n+1)^2-2r(2m+1)(2n+1)\bigr)\Bigr)}
{(2m+1)(2n+1)}.
\end{equation}
Since
\begin{align}\nonumber
(2m+1)^2+(2n+1)^2-2r(2m+1)(2n+1)&=(1-r)\bigl((2m+1)^2+(2n+1)^2\bigr)+4r(m-n)^2\\
&\geq(1-r)\bigl((2m+1)^2+(2n+1)^2\bigr),
\end{align}
we obtain the factorized bound
\begin{equation}
\frac12-\varepsilon_\lambda(r)
\leq\frac{4}{\pi^2}\biggl(
\sum_{k=0}^{\infty}
\frac{{\rm e}^{-\lambda^2(1-r)(2k+1)^2/2}}{2k+1}
\biggr)^2.
\label{eq:S-factorized-upper}
\end{equation}
Set $c=\lambda^2(1-r)/2>0$. The inequalities
\begin{equation}
(2k+1)^2\geq1+8k,\qquad\frac{1}{2k+1}\leq1,
\qquad k=0,1,\ldots,
\end{equation}
give
\begin{equation}
\sum_{k=0}^{\infty}\frac{{\rm e}^{-c(2k+1)^2}}{2k+1}
\leq{\rm e}^{-c}\sum_{k=0}^{\infty}{\rm e}^{-8ck}
=\frac{{\rm e}^{-c}}{1-{\rm e}^{-8c}}.
\label{eq:S-odd-sum-bound}
\end{equation}
Substituting this into Eq.~\eqref{eq:S-factorized-upper} yields
\begin{equation}
\frac12-\varepsilon_\lambda(r)
\leq\frac{4}{\pi^2}
\frac{{\rm e}^{-\lambda^2(1-r)}}
{\bigl(1-{\rm e}^{-4\lambda^2(1-r)}\bigr)^2},
\end{equation}
which completes the proof of Eq.~\eqref{eq:S-large-lambda}. For each
fixed $r<1$, the upper bound tends to zero as $\lambda\to\infty$,
proving Eq.~\eqref{eq:S-large-lambda-limit}.

If in addition $r>0$, division of Eq.~\eqref{eq:S-large-lambda}
by the common leading term gives
\begin{equation}
1-{\rm e}^{-2\lambda^2r}
\leq
\frac{\frac12-\varepsilon_\lambda(r)}
{\frac{4}{\pi^2}{\rm e}^{-\lambda^2(1-r)}}
\leq\frac{1}{\bigl(1-{\rm e}^{-4\lambda^2(1-r)}\bigr)^2}.
\end{equation}
Both bounds converge to one, which proves
Eq.~\eqref{eq:S-large-lambda-sharp}.

Finally, when $r=0$, $U$ and $V$ are independent and the square-wave
outputs are unbiased, so $\varepsilon_\lambda(0)=1/2$. When $r=1$,
$U=V$ almost surely and $\varepsilon_\lambda(1)=0$. More generally,
$\varepsilon_\lambda(r)\to0$ as $r\to1$ at each fixed $\lambda$:
under the coupling~\eqref{eq:S-UV-definition}, $V\to U$ almost
surely, and the classifier is continuous at $U$ with probability one.
Thus the limit~\eqref{eq:S-large-lambda-limit} cannot be uniform
up to $r=1$.
\end{proof}

\subsection{Linear response at small load}
Differentiating Eq.~\eqref{eq:S-p} at $\mu=0$, $v=1$,
\begin{equation}
\kappa_\lambda\coloneqq\partial_\mu p_\lambda(\mu,v)|_{\mu=0,v=1}=\frac{1}{2}\E_z\big[z\,s_\lambda(z)\big]=\frac{2\lambda}{\pi}\sum_{k\geq0}{\rm e}^{-(2k+1)^2\lambda^2/2}=\sum_{m\in\Z}(-1)^m\varphi\bigl(\frac{m\pi}{\lambda}\bigr),
\label{eq:S-kappa}
\end{equation}
where the second equality is by Stein's Lemma and the last by Poisson summation. From $\mathcal{J}_\lambda(0^+)=\kappa_\lambda^2$ and Eq.~\eqref{eq:S-stationary}, $\alpha_\lambda(r)=r/(4\kappa_\lambda^2)+O(r^2)$, hence
\begin{equation}
r_\star=4\kappa_\lambda^2\,\alpha+O(\alpha^2),\qquad \phi_{\rm RS}=-\alpha\ln2+4\kappa_\lambda^4\alpha^2+O(\alpha^3).
\end{equation}
The same value follows exactly for $P=O(1)$ patterns: the posterior mean is $\langle\bw\rangle\simeq\sum_\mu\bx^\mu\E[h|s_\lambda(h)=\xi^\mu]/\sqrt N$ and $\E[h\,|\,s_\lambda(h)=\pm1]=\pm2\kappa_\lambda$. For $\lambda\to0^+$, $4\kappa_\lambda^2\to2/\pi$; for $\lambda\gg1$, $4\kappa_\lambda^2=(16/\pi^2)\lambda^2{\rm e}^{-\lambda^2}[1+O({\rm e}^{-4\lambda^2})]$. At $\lambda=3$ and $\alpha=4$ this gives an overlap value of $\simeq0.00717$, in agreement with the numerical maximizer of the weakly aligned branch.  At this load that branch is metastable: the equilibrium maximizer $r_\star$ instead lies on the
strongly aligned branch.

\subsection{Upper spinodal}

On the weakly aligned branch $r=O(\lambda^{-2})$. For $r\le\bar r<1$ and $\lambda\to\infty$, Eq.~\eqref{eq:S-p} gives $p_\lambda-\frac{1}{2}=\frac2\pi\sin\theta\,{\rm e}^{-\lambda^2(1-r)/2}(1+O({\rm e}^{-4\lambda^2(1-r)}))$ with $\theta=\lambda\sqrt r\,t$, so $p(1-p)=\frac14+O({\rm e}^{-\lambda^2(1-r)})$ and, from Eq.~\eqref{eq:S-J} with $\E_{t\sim\mathcal{N}(0,1)}\cos^2(\lambda\sqrt r\,t)=\frac{1}{2}(1+{\rm e}^{-2\lambda^2r})$,
\begin{equation}
\mathcal{J}_\lambda(r)=\frac{4\lambda^2}{\pi^2}\,{\rm e}^{-\lambda^2}\cosh(\lambda^2r)\bigl(1+O({\rm e}^{-\lambda^2(1-r)})\bigr),\qquad
\alpha_\lambda(r)=\frac{\pi^2}{16}\,\frac{{\rm e}^{\lambda^2}}{\lambda^4}\,\frac{x}{(1-x/\lambda^2)\cosh x}\bigl(1+O({\rm e}^{-\lambda^2(1-r)})\bigr),
\qquad x=\lambda^2r.
\end{equation}
The first fold (local maximum of $\alpha_\lambda$) is at $x=x_0+O(\lambda^{-2})$ with $x_0\tanh x_0=1$, $x_0=1.19968$, which gives
\begin{equation}
\alpha_\spin^+(\lambda)=C_+\,\frac{{\rm e}^{\lambda^2}}{\lambda^4}\bigl(1+O(\lambda^{-2})\bigr),\qquad C_+=\frac{\pi^2}{16}\frac{x_0}{\cosh x_0}=0.40881.
\label{eq:S-asp}
\end{equation}

\subsection{Aligned branch: boundary layers and universal error}

For fixed $\lambda$ and $r\to1$, the smoothing width $\lambda\sqrt{1-r}$ of $p_\lambda$ vanishes. As a function of $t$, $\psi_\lambda(r,t)$ is then exponentially close to $0$ or $1$ except in layers of width $w=\sqrt{(1-r)/r}$ around the fringes $t_k=k\pi/(\lambda\sqrt r)$, where $\psi_\lambda\simeq\Phi(\pm(t-t_k)/w)$ with alternating signs. Summing the layers with weights $\varphi(t_k)$,
\begin{gather}
\E_{t\sim\mathcal{N}(0,1)}H_2(\psi_\lambda)=C_H\,\Theta_\lambda\,w\,(1+o(1)),\qquad \varepsilon_\lambda(r)=\frac{2}{\sqrt\pi}\,\Theta_\lambda\,w\,(1+o(1)),\nonumber\\
C_H=\int_\R H_2[\Phi(x)]\,\de x=1.80639,\qquad \Theta_\lambda=\sum_{k\in\Z}\varphi\bigl(\frac{k\pi}{\lambda}\bigr),
\label{eq:S-layers}
\end{gather}
where we used $\int_\R2\Phi(1-\Phi)\,\de x=2/\sqrt\pi$. Note that $\Theta_\lambda\to\varphi(0)$ for $\lambda\to0^+$ (a single boundary) and $\Theta_\lambda=\frac\lambda\pi(1+2\sum_{j\geq1}{\rm e}^{-2j^2\lambda^2})$ by Poisson summation. With $\frac{1}{2}\ln(1-r)+\frac{r}{2}=\ln w+\frac{1}{2}+o(1)$, the reduced potential becomes $\mathcal{F}_\lambda\simeq\ln w+\frac{1}{2}-\alpha C_H\Theta_\lambda w$, maximized at $w_\star=1/(\alpha C_H\Theta_\lambda)$:
\begin{equation}
\mathcal{F}^{(+)}_\lambda(\alpha)\simeq-\ln(\alpha\,C_H\,\Theta_\lambda)-\frac{1}{2},\qquad
1-r_\star\simeq\frac{1}{(\alpha C_H\Theta_\lambda)^2},\qquad
\varepsilon_g\simeq\frac{2}{\sqrt\pi\,C_H}\,\frac1\alpha=\frac{0.62466}{\alpha}.
\label{eq:S-aligned}
\end{equation}
The approximation requires $w_\star$ to be small compared with the fringe spacing, $\alpha C_H\Theta_\lambda\gg\max(1,\lambda/\pi)$. The error is $\lambda$ independent and coincides with the classical Gibbs-learning law of the spherical perceptron~\cite{Seung1992,engel2001}. As a check, at $\lambda=3$, $\alpha=4$ Eq.~\eqref{eq:S-aligned} gives $\mathcal{F}^{(+)}\simeq-2.4315$ versus the exact $-2.44171$, and $\alpha\varepsilon_g$ at $\alpha=50$ equals $0.6243$, $0.6245$ and $0.6246$ for $\lambda\to0^+$, $2$ and $3$.

\subsection{Coexistence line at large frequency}

For $\lambda\gg1$, $\mathcal{F}^{(-)}_\lambda=-\alpha\ln2+O(\alpha^2\lambda^4{\rm e}^{-2\lambda^2})$ and $\Theta_\lambda\simeq\lambda/\pi$. Equating with Eq.~\eqref{eq:S-aligned},
\begin{equation}
\alpha_{\rm eq}\ln2-\ln\alpha_{\rm eq}=\ln\lambda+c_0+o(1),\qquad c_0=\ln\frac{C_H}{\pi}+\frac{1}{2}=-0.05340,
\label{eq:S-aeq}
\end{equation}
so that $\alpha_{\rm eq}=\log_2\lambda+\log_2\log_2\lambda+O(1)$. Table~\ref{tab:asym} compares Eqs.~\eqref{eq:S-asp} and~\eqref{eq:S-aeq} with the numerical solution.

\subsection{Lower spinodal at large frequency}

At fixed $s=\lambda^2(1-r)$ and $\lambda\to\infty$, the phase $\theta=\lambda\sqrt r\,t$ is equidistributed modulo $2\pi$, and $\mathcal{F}_\lambda=-\ln\lambda+\frac{1}{2}+\frac{1}{2}\ln s-\alpha\ln 2+\alpha D(s)+o(1)$ with
\begin{equation}
D(s)=\ln2-\frac{1}{2\pi}\int_0^{2\pi}H_2\big[\bar p(\theta,s)\big]\,\de\theta,\qquad \bar p(\theta,s)=\mathbb P_z\big[\sin(\theta+\sqrt s\,z)>0\big],
\end{equation}
the information carried by one label in the aligned phase. The aligned branch is stationary at $\alpha=-1/[2sD'(s)]$, whose minimum over $s$ gives $\alpha_\spin^-(\infty)=2.6466$ (at $s=0.713$). The finite-$\lambda$ values $\alpha_\spin^-=2.422$, $2.485$, $2.524$, $2.569$ at $\lambda=3$, $3.5$, $4$, $5$ approach it from below.

\begin{table}[t]
\caption{Large-frequency asymptotics versus the numerical RS solution. $\alpha_{\rm eq}^{\rm asy}$ solves Eq.~\eqref{eq:S-aeq} without the $o(1)$ term; $\alpha_\spin^{+,\rm asy}=C_+{\rm e}^{\lambda^2}/\lambda^4$.}
\label{tab:asym}
\begin{ruledtabular}
\begin{tabular}{cccccc}
$\lambda$ & $\alpha_{\rm eq}$ & $\alpha_{\rm eq}^{\rm asy}$ & $\alpha_\spin^+$ & $\alpha_\spin^+/\alpha_\spin^{+,\rm asy}$ & $\alpha_\spin^-$\\
\hline
3.0 & 3.2125 & 3.1744 & 47.61 & 1.164 & 2.4216\\
3.5 & 3.5855 & 3.5637 & 633.9 & 1.113 & 2.4848\\
4.0 & 3.8922 & 3.8784 & $1.538\times10^{4}$ & 1.084 & 2.5243\\
5.0 & 4.3805 & 4.3738 & $4.952\times10^{7}$ & 1.051 & 2.5694\\
\end{tabular}
\end{ruledtabular}
\end{table}

%=====================================================================
\section{Structure of the reduced potential}\label{sec:structure}
%=====================================================================

Here, we investigate the structure of the reduced potential, showing that it satisfies a heat equation and admits a Fisher-information representation. We further derive the conditions for the occurrence of first-order transitions within the RS framework.

\begin{prop}[Heat equation]\label{prop:heat}
$p_\lambda$ is smooth on $\R\times(0,\infty)$ and $\partial_vp_\lambda=\frac{1}{2}\partial_\mu^2p_\lambda$. Consequently, for $0<r<1$,
\begin{equation}
2r\,\partial_r\psi_\lambda(r,t)=t\,\partial_t\psi_\lambda(r,t)-\partial_t^2\psi_\lambda(r,t).
\label{eq:S-OU}
\end{equation}
\end{prop}
\begin{proof}
$p_\lambda(\mu,v)=\int\de x\,G_v(x-\mu)\Theta(\sin\lambda x)$ with $G_v$ the heat kernel, which satisfies $\partial_vG_v=\frac{1}{2}\partial_x^2G_v$; derivatives can be taken on the kernel because $\Theta(\sin\lambda x)$ is bounded. With $\psi_\lambda=p_\lambda(\sqrt r\,t,1-r)$: $\partial_r\psi_\lambda=\frac{t}{2\sqrt r}\partial_\mu p_\lambda-\partial_vp_\lambda=\frac{t}{2\sqrt r}\partial_\mu p_\lambda-\frac{1}{2}\partial^2_\mu p_\lambda$, $\partial_t\psi_\lambda=\sqrt r\,\partial_\mu p_\lambda$, $\partial_t^2\psi_\lambda=r\,\partial_\mu^2p_\lambda$.
\end{proof}

\begin{prop}[Fisher-information representation]\label{prop:J}
For $0<r<1$,
\begin{equation}
\mathcal{J}_\lambda(r)\coloneqq\mathcal{I}_\lambda'(r)=\frac{1}{4r}\int\Dt\,\frac{(\partial_t\psi_\lambda(r,t))^2}{\psi_\lambda(1-\psi_\lambda)}
=\frac{\lambda^2}{4}\,\E_{t\sim\mathcal{N}(0,1)}\bigg[\frac{(\partial_\theta p)^2}{p(1-p)}\bigg]_{p=p_\lambda(\theta/\lambda,1-r),\ \theta=\lambda\sqrt r\,t}>0 .
\label{eq:S-J}
\end{equation}
In particular $\mathcal{I}_\lambda$ is strictly increasing and $\E_{t\sim\mathcal{N}(0,1)}H_2(\psi_\lambda(r,t))$ strictly decreasing in $r$. Moreover $\mathcal{J}_\lambda(0^+)=\kappa_\lambda^2$ with $\kappa_\lambda$ given in Eq.~\eqref{eq:S-kappa}.
\end{prop}
\begin{proof}
Since $\int\Dt\,\partial_r\psi_\lambda=\partial_r\frac{1}{2}=0$, $\mathcal{I}_\lambda'=\int\Dt\,\partial_r\psi_\lambda\ln\psi_\lambda$. Insert Eq.~\eqref{eq:S-OU} and use Gaussian integration by parts in the form $\int\Dt\,(t f'-f'')g=\int\Dt\,f'g'$ with $f=\psi_\lambda$, $g=\ln\psi_\lambda$ (boundary terms vanish because $\psi_\lambda$ and its derivatives are bounded away from the singularities of the logarithm for $r<1$): $\mathcal{I}_\lambda'=\frac{1}{2r}\int\Dt\,(\partial_t\psi_\lambda)^2/\psi_\lambda$. The symmetry $\psi_\lambda(r,-t)=1-\psi_\lambda(r,t)$ implies that $(\partial_t\psi_\lambda)^2$ is even, so the same integral equals $\frac{1}{2r}\int\Dt\,(\partial_t\psi_\lambda)^2/(1-\psi_\lambda)$; averaging the two forms gives Eq.~\eqref{eq:S-J}. The second form follows from $\partial_t=\lambda\sqrt r\,\partial_\theta$. The small-$r$ limit follows from $\partial_t\psi_\lambda\to\sqrt r\,\partial_\mu p_\lambda(0,1)=\sqrt r\,\kappa_\lambda$ and $\psi_\lambda\to1/2$.
\end{proof}

Since $\partial_r\frac{1}{2}(\ln(1-r)+r)=-r/(2(1-r))$,
\begin{equation}
\partial_r\mathcal{F}_\lambda(r;\alpha)=2\mathcal{J}_\lambda(r)\bigl(\alpha-\alpha_\lambda(r)\bigr),\qquad \alpha_\lambda(r)=\frac{r}{4(1-r)\mathcal{J}_\lambda(r)},\qquad
\partial_r^2\mathcal{F}_\lambda\big|_{\alpha=\alpha_\lambda(r)}=-2\mathcal{J}_\lambda(r)\,\alpha_\lambda'(r).
\label{eq:S-stationary}
\end{equation}
Every interior stationary point lies on the curve $\alpha=\alpha_\lambda(r)$; increasing (decreasing) portions are local maxima (minima). Since $\partial_r\mathcal{F}_\lambda(0^+)=2\alpha\kappa_\lambda^2>0$, the boundary $r=0$ is never a maximizer for $\alpha>0$. A critical endpoint of a line of first-order transitions occurs where the two folds $\alpha_\lambda'(r)=0$ merge, $\alpha_\lambda'=\alpha_\lambda''=0$.

\begin{thm}[First-order transition in the RS theory]\label{thm:first-order}
Fix $\lambda>0$ and suppose that at $\alpha_{\rm eq}$ the function $\mathcal{F}_\lambda(\cdot;\alpha_{\rm eq})$ has exactly two global maximizers $0<r_-<r_+<1$, both nondegenerate ($\partial_r^2\mathcal{F}_\lambda<0$). Then the two maxima continue smoothly as $r_\pm(\alpha)$ near $\alpha_{\rm eq}$, the low (high) overlap maximum is the global one for $\alpha<\alpha_{\rm eq}$ ($\alpha>\alpha_{\rm eq}$) close to $\alpha_{\rm eq}$, the equilibrium overlap jumps by $r_+-r_->0$, and
\begin{equation}
\partial_\alpha\phi_{\rm RS}\big|_{\alpha_{\rm eq}^+}-\partial_\alpha\phi_{\rm RS}\big|_{\alpha_{\rm eq}^-}=2\bigl(\mathcal{I}_\lambda(r_+)-\mathcal{I}_\lambda(r_-)\bigr) =\E_{t\sim\mathcal{N}(0,1)}H_2\bigl(\psi_\lambda(r_-,t)\bigr) - \E_{t\sim\mathcal{N}(0,1)}H_2\bigl(\psi_\lambda(r_+,t)\bigr)>0,
\label{eq:S-jump}
\end{equation}
while $\phi_{\rm RS}$ is continuous.
\end{thm}
\begin{proof}
The implicit function theorem gives smooth branches $r_\pm(\alpha)$. Let $\Delta\mathcal{F}(\alpha)=\mathcal{F}_\lambda(r_+(\alpha);\alpha)-\mathcal{F}_\lambda(r_-(\alpha);\alpha)$. By stationarity, $\Delta\mathcal{F}'(\alpha)=\partial_\alpha\mathcal{F}_\lambda(r_+;\alpha)-\partial_\alpha\mathcal{F}_\lambda(r_-;\alpha)=2(\mathcal{I}_\lambda(r_+)-\mathcal{I}_\lambda(r_-))>0$ by Proposition~\ref{prop:J}. Hence $\Delta\mathcal{F}<0$ just below and $>0$ just above $\alpha_{\rm eq}$; since no other maximizer is global at $\alpha_{\rm eq}$, continuity excludes other branches nearby. Along each smooth branch $\frac{\de}{\de\alpha}\phi_{\rm RS}=\partial_\alpha\mathcal{F}_\lambda=2\mathcal{I}_\lambda(r_\star)$, which gives Eq.~\eqref{eq:S-jump}.
\end{proof}
By Eq.~\eqref{eq:S-predictive}, Eq.~\eqref{eq:S-jump} is a discontinuous drop of the predictive entropy of a fresh label. Stability here refers to the scalar RS potential; it is not a stability analysis against replica-symmetry breaking.

%=====================================================================
\section{Message passing}\label{sec:GAMP}
%=====================================================================

We discuss here the Bayes-optimal generalized approximate
message passing (GAMP) method ~\cite{Donoho2009,Rangan2011,Zdeborova2016} adapting it to our setting.

The GAMP algorithm relies on a factorized weight prior, an output likelihood, and a dense random input matrix with independent, centered entries of order $N^{-1/2}$. In particular, with this last assumption, each local field --- defined from the random input matrix --- is a sum of many weak contributions and, within the message-passing approximation, admits a Gaussian approximation at large $N$, reducing the output-message calculations to one-dimensional integrals.

For the case considered in this work, the random input matrix is  $F_{\mu i}=x_i^\mu/\sqrt N$, and the output channel is
$P_{\rm out}(\xi\mid h)=\one\{s_\lambda(h)=\xi\}$.
\begin{remark}
Note that the GAMP algorithm uses independent $\mathcal{N}(0,1)$ weight priors. This differs from the spherical prior of our model, at finite $N$. Hence, our comparison rests on their common reduced RS potential, as explained in Remark~\ref{rem.dist_lim} of Appendix~\ref{sec:nishimori}. In particular, the state evolution we derive below is the matched large-system description used in this comparison. Its application to our Rademacher inputs and discontinuous channel carries the qualifications discussed in Appendices~\ref{sec:replica} and~\ref{sec:nishimori}.
\end{remark}

For a Gaussian field prediction of mean $\omega$ and variance $V$, define
\begin{equation} \label{eq:gamp-output}
 Z_\xi(\omega,V)\coloneqq \int dh\,\mathcal{N}(h;\omega,V)
             P_{\rm out}(\xi\mid h),
 \qquad
 g_{\rm out}(\omega,V,\xi)\coloneqq \partial_\omega\ln Z_\xi(\omega,V).
\end{equation}
In the model we consider, $Z_{+1}=p_\lambda(\omega,V)$ and
$Z_{-1}=1-p_\lambda(\omega,V)$, with $p_\lambda$ defined in~\eqref{eq:S-p}.
Thus, $Z_\xi$ is the probability assigned to the observed label under the
Gaussian field prediction. The quantity $Vg_{\rm out}$ is the change in its
mean after conditioning on that label. Let $\widehat{w}_i^k$ and $v_i^k$ be the estimated mean and variance of weight
$i$ at iteration $k$, respectively. The undamped updates then give
\begin{equation}\label{eq:gamp-fields}
 V_\mu^k=\sum_i F_{\mu i}^2v_i^k,
 \qquad
 \omega_\mu^k=\sum_iF_{\mu i}\widehat w_i^k-V_\mu^k g_\mu^{k-1},
\end{equation}
where the term subtracted in $\omega_\mu^k$ takes into account the so-called \emph{Onsager correction}. After the update, we can then recompute $g_\mu^k=g_{\rm out}(\omega_\mu^k,V_\mu^k,\xi^\mu)$ and $a_\mu^k=-\partial_\omega g_{\rm out}(\omega_\mu^k,V_\mu^k,\xi^\mu)$.
The observation messages yield a quadratic approximation to the effective log-likelihood of each weight, with $B_i^k$ and $A_i^k$ controlling its linear and quadratic terms, respectively. Combining it with the standard Gaussian prior gives
\begin{equation} \label{eq:gamp-weights}
 \begin{aligned}
 A_i^k&=\sum_\mu F_{\mu i}^2a_\mu^k,
 & B_i^k&=\sum_\mu F_{\mu i}g_\mu^k,\\
 \widehat w_i^{k+1}
   &=\frac{A_i^k\widehat w_i^k+B_i^k}{1+A_i^k},
 & v_i^{k+1}&=\frac{1}{1+A_i^k}.
 \end{aligned}
\end{equation}
The uninformed initialization corresponds to $\widehat{\bm w}^{\,0}=0$,
$v_i^0=1$, and $g_\mu^{-1}=0$. The finite-$N$ implementation damps these
updates; numerical details are given in Appendix~\ref{sec:numerics}.

State evolution describes the typical behavior of these vector updates as
$N,P\to\infty$ at fixed $\alpha=P/N$. On the matched Bayes-optimal
trajectory, the limiting teacher overlap and squared norm agree, i.e., 
\begin{equation} \label{eq:gamp-order-parameters}
 m_k=\lim_{N\to\infty}\frac{\widehat{\bm w}^{\,k}\!\cdot\bm w^*}{N},
 \qquad
 q_k=\lim_{N\to\infty}\frac{\|\widehat{\bm w}^{\,k}\|^2}{N},
 \qquad m_k=q_k=:r_k.
\end{equation}
The associated mean squared error per weight is $1-r_k$.
The effective teacher field conditional on its prediction is
$h^*\mid\omega\sim\mathcal{N}(\omega,1-r_k)$, with
$\omega=\sqrt{r_k}\,t$ and $t\sim\mathcal{N}(0,1)$. Hence $\mathbb{P}(\xi=+1\mid t)=\psi_\lambda(r_k,t)$.
Writing $p=p_\lambda(\omega,V)$, averaging the two possible label scores
gives $\E_{\xi\mid\omega}[g_{\rm out}^2]
=(\partial_\omega p)^2/[p(1-p)]$, the same value being achieved also by the average curvature
$\E_{\xi\mid\omega}[a]$. Since $F_{\mu i}^2=1/N$, the coefficient $A_i^k$ is independent of the weight index $i$ and converges in the large-system limit to
\begin{equation}\label{eq:gamp-precision}
 A_k=\alpha\E[g_{\rm out}^2]
 =\alpha\E_t\!\left[
 \frac{[\partial_\omega p_\lambda(\sqrt{r_k}\,t,1-r_k)]^2}
 {\psi_\lambda(r_k,t)[1-\psi_\lambda(r_k,t)]}\right]
 =4\alpha\mathcal{J}_\lambda(r_k),
\end{equation}
where the last equality is the Fisher-information identity~\eqref{eq:S-J}.

Let now $R_i^k=\widehat w_i^k+\frac{B_i^k}{A_i^k}$ denote the center of the effective Gaussian likelihood for weight $i$, obtained by combining the observation messages at iteration $k$. In the matched large-system description, the joint statistics of $w_i^*$ and $R_i^k$ are represented by the scalar model $R=W^*+\frac{Z}{\sqrt{A_k}}$,
where $W^*$ represents a typical teacher-weight component, $W^*$ and \(Z\) are independent standard Gaussian variables, and $A_k$ is the limiting value of $A_i^k$. Thus, $R$ acts as a noisy observation of the teacher weight, with noise variance $1/A_k$. Combining the corresponding Gaussian likelihood with the standard Gaussian prior gives the posterior mean and variance used in the next weight update, i.e.,
\begin{equation}\label{eq:gamp-scalar-input}
 \widehat W^{\,\mathrm{new}}
 =\frac{A_k}{1+A_k}R
 =\frac{A_k}{1+A_k}W^*+\frac{\sqrt{A_k}}{1+A_k}Z,
 \qquad
 \E[W^*\widehat W^{\,\mathrm{new}}]
 =\E[(\widehat W^{\,\mathrm{new}})^2]
 =\frac{A_k}{1+A_k}.
\end{equation}
From this, recalling the definition of $r_k$ in~\eqref{eq:gamp-order-parameters}, and taking into account~\eqref{eq:gamp-precision} and~\eqref{eq:gamp-scalar-input}, one can then derive the overlap recurrence relation 
\begin{equation}\label{eq:gamp-se}
 r_{k+1}=T_{\alpha,\lambda}(r_k)
 \coloneqq\frac{4\alpha\mathcal{J}_\lambda(r_k)}
 {1+4\alpha\mathcal{J}_\lambda(r_k)}.
\end{equation}
To connect this recurrence relation to the RS calculation, set
$S_\lambda(r)\coloneqq\E_t H_2[\psi_\lambda(r,t)]$.
The heat equation~\eqref{eq:S-OU} and Gaussian integration by parts give $S_\lambda'(r) =-\frac{1}{2r}\E_t\!\left[\frac{(\partial_t\psi_\lambda)^2}{\psi_\lambda(1-\psi_\lambda)}
 \right]=-2\mathcal{J}_\lambda(r)$, and $\partial_r\mathcal{F}_\lambda(r;\alpha)=-\frac{r}{2(1-r)}+2\alpha\mathcal{J}_\lambda(r)$.
Thus the RS saddle-point equation is precisely the fixed-point condition
for~\eqref{eq:gamp-se}, that is,
\begin{equation}
 \partial_r\mathcal{F}_\lambda(r;\alpha)=0
 \quad\Longleftrightarrow\quad
 \frac{r}{1-r}=4\alpha\mathcal{J}_\lambda(r)
 \quad\Longleftrightarrow\quad
 r=T_{\alpha,\lambda}(r).
 \label{eq:gamp-rs-fixed-points}
\end{equation}
The same self-consistency equation therefore governs the RS stationary
overlaps and the fixed points of matched GAMP state evolution.
Their comparison concerns macroscopic order parameters: GAMP estimates
posterior means, whereas the Gibbs generalization error in~\eqref{eq:eps} refers
to posterior samples. The Nishimori identities identify the averaged
teacher--mean and replica overlaps; with the assumed overlap concentration,
these yield the common scalar parameter $r$.

\begin{figure}[H]
    \centering
    \subfloat[\label{fig_gamp_r}]{%
        \includegraphics[width=0.48\columnwidth]{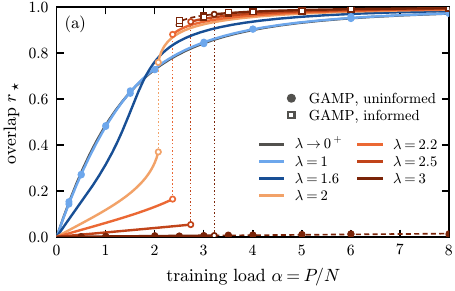}}
    \hfill
    \subfloat[\label{fig_gamp}]{%
        \includegraphics[width=0.48\columnwidth]{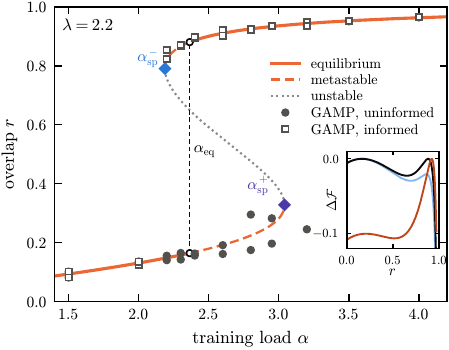}}
    \caption{(a) Equilibrium RS overlaps and GAMP estimates at $N=4000$. Filled circles and open squares denote uninformed and informed initializations, respectively. Dotted vertical segments mark equilibrium jumps; dashed curves show metastable branches for $\lambda=3$. (b) Stationary branches at $\lambda=2.2$: Equilibrium (solid orange), metastable (dashed orange), and unstable (dotted gray). Diamonds mark spinodals; the vertical dashed line marks coexistence. Inset: $\mathcal F_\lambda-\max_r\mathcal F_\lambda$ at $\alpha=2.28$ (light blue), $\alpha_{\mathrm{eq}}$ (black), and $2.70$ (red).
    \label{fig5}}
\end{figure}

The selection of a solution remains essential. RS equilibrium maximizes
$\mathcal{F}_\lambda$, while GAMP converges to an attracting fixed point
accessible from its initialization. At a fixed point, local attraction
requires $|T_{\alpha,\lambda}'(r)|<1$, and
\begin{equation}\label{eq:gamp-stability}
 T_{\alpha,\lambda}'(r)
   =4\alpha(1-r)^2\mathcal{J}_\lambda'(r),
 \qquad
\partial_r^2\mathcal{F}_\lambda(r;\alpha)
   =\frac{T_{\alpha,\lambda}'(r)-1}{2(1-r)^2}.
\end{equation}
Where $\mathcal{J}_\lambda'(r)\geq0$, a nondegenerate local maximum is
attracting and a local minimum is repelling. This includes both equilibrium
and metastable maxima in the regime illustrated in Fig.~\ref{fig_gamp}. When the state-evolution map $T_{\alpha,\lambda}$ is nondecreasing in $r$, iteration from the uninformed initialization $r_0=0$ converges to its smallest fixed point, explaining why an uninformed start follows the weak branch while that branch exists. This monotonicity is verified numerically in the investigated range $0.5\leq\lambda\leq4$. For $\lambda>\lambda_c$, the coexistence interval contains three stationary overlaps $r_{\rm w}<r_{\rm u}<r_{\rm s}$, that is, weak and strong local maxima, separated by an unstable local minimum. The strong maximum and the intermediate minimum appear together at the lower spinodal $\alpha_{\rm sp}^-$. At $\alpha_{\rm eq}$, the two maxima have equal
height and exchange their equilibrium preference. The weak maximum
survives until $\alpha_{\rm sp}^+$, where it merges with the intermediate
minimum. The expected branch selection is summarized in
Table~\ref{tab:gamp-branches}.

\begin{table}[htbp]
 \centering
 \small
 \setlength{\tabcolsep}{4pt}
 \renewcommand{\arraystretch}{1.18}
 \begin{tabularx}{\textwidth}{@{}p{0.26\textwidth}YYYY@{}}
 \toprule
 Training load & Local maxima & RS equilibrium & Uninformed GAMP
   & Informed GAMP\\
 \midrule
 $\alpha<\alpha_{\rm sp}^-$
   & Weak only & Weak & Weak & Relaxes to weak\\
 $\alpha_{\rm sp}^-<\alpha<\alpha_{\rm eq}$
   & Weak and strong & Weak & Weak & Can reach strong\\
 $\alpha_{\rm eq}<\alpha<\alpha_{\rm sp}^+$
   & Weak and strong & Strong & Remains weak & Can reach strong\\
 $\alpha>\alpha_{\rm sp}^+$
   & Strong only & Strong & Can reach strong & Strong\\
 \bottomrule
 \end{tabularx}
 \caption{Expected branch selection in the first-order regime within the
 scalar state-evolution description. An informed start is placed
 sufficiently close to the strong attractor whenever it exists. In the two
 middle intervals an intermediate unstable stationary point is also
 present. The open intervals exclude coexistence and the spinodals.
 Finite-$N$ and finite-time effects can modify the apparent switching loads.}
 \label{tab:gamp-branches}
\end{table}

The two numerical protocols --- i.e., informed and uninformed --- use the same GAMP updates. The uninformed
start supplies no information about the teacher direction. The informed
start, $\widehat{\bm w}^{\,0}=\sqrt{0.995}\,\bm w^*$ with
$v_i^0=0.005$, supplies teacher information to test the existence and
stability of the strong attractor. This diagnostic initialization need
not satisfy the matched scalar identity $m_k=q_k$ at finite iterations.
The measured finite-$N$ overlap is $m=\widehat{\bm w}\cdot\bm w^*/N$. Figure~\ref{fig:scurve} makes the table clear at $\lambda=2.2$, where $\alpha_{\rm sp}^-\simeq2.18717$, $\alpha_{\rm eq}\simeq2.36387$, and $\alpha_{\rm sp}^+\simeq3.04040$. The solid orange curve denotes the equilibrium maximum, the dashed orange
curve the metastable maximum, and the dotted gray curve the intermediate
minimum. Diamonds mark the spinodals and the vertical dashed line marks
coexistence. Filled circles from uninformed GAMP follow the weak branch
beyond $\alpha_{\rm eq}$ and leave it near $\alpha_{\rm sp}^+$; open
squares from informed GAMP follow the strong branch down to
$\alpha_{\rm sp}^-$. For example, at $\lambda=2.2$ and $\alpha=2.5$, state 
evolution has a weakly aligned fixed point at $r_{\mathrm w} \simeq 0.17960$ 
and a strongly aligned fixed point at $r_{\mathrm s} \simeq 0.90083$. The RS equilibrium selects the strongly aligned solution, whereas GAMP from an 
uninformed initialization converges to the weakly aligned one.
Thus $\alpha_{\rm eq}<\alpha<\alpha_{\rm sp}^+$ is the interval where
the equilibrium improvement is inaccessible to this uninformed iteration. Fig.~\ref{fig_gamp_r} extends this comparison to several readout frequencies. At $\lambda=1$, GAMP follows the smooth equilibrium RS curve, whereas at $\lambda=2.2$ and $3$ its behavior depends on initialization. Specifically, informed runs track the strongly aligned branch when it exists, while uninformed runs remain weakly aligned beyond coexistence. For $\lambda=3$, this persistence extends throughout the displayed load range above $\alpha_{\mathrm{eq}}$.

%=====================================================================
\section{Numerical methods and validation}\label{sec:numerics}
%=====================================================================
We describe hereafter the numerical methods used to evaluate the RS functionals, select the physical branch, and treat coexistence and critical points. We additionally present the implementation and validation of Bayes-optimal GAMP.

All averages over $t$ are computed exactly in the phase variable: $\theta=\lambda\sqrt r\,t$ modulo $2\pi$ has the wrapped-normal density with standard deviation $\lambda\sqrt r$, evaluated by direct image sums (small width) or its Fourier series (large width). Using $\psi(\theta+\pi)=1-\psi(\theta)$ and $\psi(\pi-\theta)=\psi(\theta)$, the integrands $H_2(\psi)$, $\psi(1-\psi)$ and $(\partial_\theta\psi)^2/\psi(1-\psi)$ reduce to $[0,\pi/2]$, where they are integrated by composite Gauss-Legendre quadrature (order 24) on a mesh graded geometrically toward the fringe $\theta=0$ at the scale $\min(\lambda\sqrt r,\lambda\sqrt{1-r})$. The smoothed square wave is evaluated as a sum of Gaussian interval probabilities $\sum_k\mathbb P[\theta+\sigma z\in(2k\pi,(2k+1)\pi)]$, computing each difference of normal distribution functions on the side of the tail to avoid cancellation; $1-\psi$ is computed as an independent sum. The classical limit uses $\psi=\Phi(\sqrt{r/(1-r)}\,t)$ with an analogous graded mesh.

\emph{Checks.} (a) The quadrature for $\varepsilon_\lambda$ agrees with the double series~\eqref{eq:S-eps-series} to $\le2\times10^{-16}$ for $\lambda\in\{0.5,1,2,3\}$, $r\in\{0.05,0.3,0.7,0.95\}$. (b) $\mathcal{J}_\lambda$ from Eq.~\eqref{eq:S-J} agrees with a centered finite difference of $\mathcal{I}_\lambda$ to $\le10^{-9}$ relative accuracy, including $r=0.99$. (c) A Monte Carlo estimate of Eq.~\eqref{eq:S-epsdef} with $2\times10^6$ samples agrees within its statistical error. (d) $\mathcal{J}_\lambda(0^+)=\kappa_\lambda^2$ and the classical limit are reproduced. (e) The bounds of Propositions~\ref{prop:eps-limits} and~\ref{prop:psi-inf} hold on a grid $\lambda\in[0.3,4]$, $r\in[0,0.99]$.

For each $\lambda$ we tabulate $\alpha_\lambda(r)$ on a grid in $u=-\ln(1-r)\in[10^{-6},16]$ (about 1700 points, dense near $u=0$), locate its folds by bounded optimization, and identify the increasing portions (maxima of $\mathcal{F}_\lambda$). For given $\alpha$, each maximum is obtained by solving $\alpha_\lambda(r)=\alpha$ on each increasing portion with Brent's method, and the equilibrium is the one with the largest $\mathcal{F}_\lambda$. The coexistence load is the root of $\mathcal{F}_\lambda(r_+(\alpha);\alpha)-\mathcal{F}_\lambda(r_-(\alpha);\alpha)$ in $(\alpha_\spin^-,\alpha_\spin^+)$. The critical endpoint is the root in $\lambda$ of $\min_r\partial_u\alpha_\lambda$. Results are in Table~\ref{tab:transitions}; all values agree to the displayed digits with an independent implementation based on Gaussian interval sums and direct maximization of $\mathcal{F}_\lambda$ on a grid.

\begin{table}[t]
\caption{Numerically resolved RS transitions. Subscripts $-$ and $+$ denote the one-sided limits at $\alpha_{\rm eq}$. The critical endpoint is $(\lambda_c,\alpha_c,r_c)=(1.94359,1.98100,0.58191)$ with $\varepsilon_\lambda(r_c)=0.41750$.}
\label{tab:transitions}
\begin{ruledtabular}
\begin{tabular}{cccccccc}
$\lambda$ & $\alpha_\spin^-$ & $\alpha_{\rm eq}$ & $\alpha_\spin^+$ & $r_-$ & $r_+$ & $\varepsilon_-$ & $\varepsilon_+$\\
\hline
2.0 & 2.04647 & 2.07129 & 2.11561 & 0.36962 & 0.75924 & 0.46913 & 0.34563\\
2.1 & 2.12698 & 2.22270 & 2.48403 & 0.24015 & 0.84018 & 0.48750 & 0.29974\\
2.2 & 2.18717 & 2.36387 & 3.04040 & 0.16441 & 0.88076 & 0.49434 & 0.27222\\
2.3 & 2.23534 & 2.49550 & 3.85776 & 0.11380 & 0.90606 & 0.49739 & 0.25293\\
2.5 & 2.30892 & 2.73308 & 6.84587 & 0.05349 & 0.93615 & 0.49947 & 0.22683\\
3.0 & 2.42162 & 3.21247 & 47.6067 & 0.00576 & 0.96842 & 0.49999 & 0.19147\\
\end{tabular}
\end{ruledtabular}
\end{table}

We run Bayes-optimal GAMP~\cite{Rangan2011,Zdeborova2016} with Gaussian prior, Rademacher inputs $F_{\mu i}=x^\mu_i/\sqrt N$, and output function $g_{\rm out}(\omega,V,y)=\partial_\omega\ln Z_y$, $Z_{+1}=p_\lambda(\omega,V)$, $Z_{-1}=1-p_\lambda(\omega,V)$, evaluated with Gaussian interval sums together with $\partial_\omega g_{\rm out}$. Iterates are damped (factor $0.6$), and runs stop when the rms change of the estimate is below $2\times10^{-6}$ or after 800 iterations. The uninformed initialization is $\hat\bw=0$, unit variances; the informed one is $\hat\bw=\sqrt{0.995}\,\bw^*$ with variances $0.005$. We use $N=4000$, $\lambda\in\{1,2.2,3\}$, and two independent instances per point; the overlap is $m=\hat\bw\cdot\bw^*/N$. Near the spinodals convergence is slow (critical slowing down) and the displayed values carry larger finite-time errors.

%=====================================================================
\section{Fragmentation of the version space}\label{sec:fragmentation}
%=====================================================================

In this last Section, we examine how the geometry of the version space evolves as the frequency parameter is varied.

For each pattern let $k_\mu\in\Z$ be the fringe index of the student field, $h_\mu\in(k_\mu\pi/\lambda,(k_\mu+1)\pi/\lambda)$; the label is $(-1)^{k_\mu}$. The version space is the disjoint union
\begin{equation}
\mathcal{V}=\bigsqcup_{\mathbf k:\,(-1)^{k_\mu}=\xi^\mu}\mathcal{C}_{\mathbf k},\qquad \mathcal{C}_{\mathbf k}=\Big\{\bw:\ \frac{\bw\cdot\bx^\mu}{\sqrt N}\in\Bigl(\frac{k_\mu\pi}{\lambda},\frac{(k_\mu+1)\pi}{\lambda}\Bigr)\ \forall\mu\Big\}.
\end{equation}
A continuous path in $\mathcal{V}$ cannot change any $k_\mu$, because $h_\mu$ would have to cross a fringe of opposite label; hence every connected component of $\mathcal{V}$ lies in a single cell. In the classical regime $\lambda<\pi/\sqrt N$, where $|h_\mu|\le\sqrt N<\pi/\lambda$, only $k_\mu\in\{0,-1\}$ occur and $\mathcal{V}$ is the intersection of the sphere with a convex cone. Any dynamics that moves continuously at zero training error (gradient flows, Langevin or billiard dynamics restricted to $\mathcal{V}$) is confined to its initial cell.

This has a practical consequence for finite-size checks. By the Nishimori property the teacher is a typical posterior sample, but a sampler started at the teacher and moving continuously samples only the teacher's cell. We compared, at $N=60$, $\alpha=0.3$, $\lambda=2.2$ (ten instances), an exact great-circle hit-and-run sampler, which jumps between cells, with a reflective (billiard) sampler that is exact within a cell. The former gives an average overlap $\langle r\rangle$ consistent with the RS/linear-response value $4\kappa_\lambda^2\alpha=0.019$ up to finite-size fluctuations, whereas the latter overestimates it by an order of magnitude (see Table~\ref{tab:cells}). Local samplers initialized at the teacher are therefore not suitable to test the equilibrium predictions; GAMP, which does not move within $\mathcal{V}$, is used instead.

\begin{table}[H]
\caption{Average teacher-student overlap at $N=60$, $P=18$ ($\alpha=0.3$), $\lambda=2.2$, over ten instances (mean $\pm$ standard error).}
\label{tab:cells}
\begin{ruledtabular}
\begin{tabular}{ccc}
cross-cell hit-and-run (exact) & teacher-cell billiard & linear response $4\kappa_\lambda^2\alpha$, $N\to\infty$\\
\hline
$0.009\pm0.010$ & $0.250\pm0.014$ & $0.0186$\\
\end{tabular}
\end{ruledtabular}
\end{table}

\twocolumngrid

\bibliography{refs}

\end{document}